\documentclass{aa}
\usepackage{txfonts}
\usepackage{graphicx}
\usepackage{aalongtable}
\begin{document}

\title{Deep imaging survey of the environment of $\alpha$ Centauri}
\subtitle{I. Adaptive optics imaging of $\alpha$\,Cen\,B with VLT-NACO}
\titlerunning{Adaptive optics imaging of the environment of $\alpha$\,Cen\,ÊB}
\authorrunning{P. Kervella et al.}
\author{
P.~Kervella\inst{1}
\and
F. Th\'evenin\inst{2}
\and
V. Coud\'e du Foresto\inst{1}
\and
F. Mignard\inst{2}}
\offprints{P. Kervella}
\mail{Pierre.Kervella@obspm.fr}

\institute{LESIA, UMR 8109, Observatoire de Paris-Meudon, 5, place Jules Janssen, 
F-92195 Meudon Cedex, France
\and D\'epartement Cassiop\'ee, UMR 6202, Observatoire de la C\^ote 
   d'Azur, BP 4229, F-06304 Nice Cedex 4, France
}
\date{Received ; Accepted}
\abstract
{$\alpha$\,Centauri is our closest stellar neighbor, at a distance of only 1.3\,pc,
and its two main components have spectral types comparable to the Sun. This is
therefore a favorable target for an imaging search for extrasolar planets. Moreover,
indications exist that the gravitational mass of $\alpha$\,Cen\,B is higher than
its modeled mass, the difference being consistent with a substellar companion
of a few tens of Jupiter masses.}
{We searched for faint comoving companions to $\alpha$\,Cen\,B. As a secondary objective,
we built a catalogue of the detected background sources.}
{We used the NACO adaptive optics system of the VLT in the $J$, $H$, and $K_s$ bands
to search for companions to $\alpha$\,Cen\,B. This instrument allowed us to
achieve a very high sensitivity to point-like sources, with a limiting magnitude of
$m_{Ks} \approx 18$ at 7" from the star. We complemented this data set with archival
coronagraphic images from the HST-ACS instrument
to obtain an accurate astrometric calibration.}
{Over the observed area, we did not detect any comoving companion to $\alpha$\,Cen\,B
down to an absolute magnitude of 19-20 in the $H$ and $K_s$ bands. 
However, we present a catalogue of 252 background objects within about 15" of the star.
This catalogue fills part of the large void area that surrounds $\alpha$\,Cen in sky surveys
due to the strong diffused light.
We also present a model of the diffused light as a function of angular distance for the NACO
instrument, that can be used to predict the background level for bright star observations.}
{According to recent numerical models, the limiting magnitude of our search
sets the maximum mass of possible companions to 20-30 times Jupiter, between
7 and 20\,AU from $\alpha$\,Cen\,B.}

\maketitle
%
\section{Introduction}

Our closest stellar neighbor, the $\alpha$\,Cen visual triple star ($d = 1.34\,$pc),
is an extremely attractive target for an extra-solar planet search. The main
components, $\alpha$\,Cen\,A (\object{HD 128620}) and B (\object{HD 128621}),
are G2V and K1V solar-like stars,
(e.g. have solar-like asteroseismic oscillation frequencies),
while the third member is the red dwarf {\it Proxima} (M5.5V).
In all imaging planet searches, the main difficulty is in retrieving the
planetary signal in the bright point-spread function (hereafter PSF)
from the star. The proximity of $\alpha$\,Cen is a clear advantage as it
allows a faint companion to be easily separated angularly from the star
itself down to orbital distances as close as a few astronomical units.
After a discussion of the potential for companions around $\alpha$\,Cen
(Sect.~\ref{pot}), we present our adaptive optics observations (Sect.~\ref{obs})
and the existing data from the HST archive (Sect.~\ref{hst_archive}),
followed by the catalogue of the detected sources (Sect.~\ref{cat})
and a discussion (Sect.~\ref{discussion}).

\section{Why search for companions around $\alpha$\,Cen ?}
\label{pot}

Two factors led us to consider the possibility of a planetary mass companion
orbiting around $\alpha$\,Cen\,B:
the mass discrepancy between models and the dynamical mass of B on one hand,
and the existence of chaotically stable orbits at intermediate distances from the star
on the other hand.

\subsection{The mass of $\alpha$\,Cen\,B}

Th\'evenin et al.~(\cite{thevenin02}, hereafter T02) have proposed a model of $\alpha$\,Cen\,B 
that reproduces well both the asteroseismic observables and the high-precision
radius measurement obtained using long-baseline interferometry
(Kervella et al.~\cite{kervella03}; Bigot et al.~\cite{bigot06}).
This model yields a mass of $M_B = 0.907 \pm 0.006 M_\odot$ for
$\alpha$\,Cen\,B, in agreement with the study by Guenther \& Demarque~(\cite{guenther00}).
Simultaneously, Pourbaix et al.~(\cite{pourbaix02}, hereafter P02)
have measured the radial velocity of
this star with an overall precision of a few m.s$^{-1}$ and deduced a
dynamical mass of $M_B = 0.934 \pm 0.006\,M_\odot$.
The difference between the model mass and the dynamical mass of B
reaches $\Delta M_B = 0.027 \pm 0.008\,M_\odot = 28 \pm 9$ Jupiter masses
(hereafter $M_J$).
No such difference is observed for $\alpha$\,Cen\,A, for which the agreement 
is excellent between the measured properties of $\alpha$\,Cen\,A and
the model from T02, computed using the mass measured by P02.

In order to explain the 3$\sigma$ difference between the modeled and measured
mass of B, a possible scenario is that the implicit assumption made by P02
of a two-body system is incorrect due to the existence of
a companion orbiting $\alpha$\,Cen B. We note that these authors
have introduced a correction of the radial velocity
of B, as they find an offset with respect to the data obtained by 
Kamper \& Wesselink~(\cite{kamper78}). This correction may mask the signature
of a long-period, low-mass companion orbiting B. The contribution
of {\it Proxima} to the radial velocity of the main pair is negligible (due to its large distance
from the A-B pair). Alternatively, a companion could also orbit
the A-B pair on a very long period orbit and currently be
located closer to B. Its gravitational contribution would make
B appear heavier in the A-B interaction. This is however less
probable, as the mass of this companion would have to be significantly higher
than the proposed $\approx 30\,M_J$ to compensate for its increased distance.

In summary, an $M \simeq 30\,M_J$ brown dwarf (hereafter BD) within
$\simeq 10$" from $\alpha$\,Cen\,B (or up to
50-100" if orbiting around the pair) could be a viable explanation
for the mass discrepancy between P02 and T02.
This hypothesis is also favored by the fact that
the $\alpha$\,Cen system is metal-rich and $\alpha$\,Cen~A
is Li-poor, as expected for stars hosting massive planets
(Santos et al.~\cite{santos03}; Israelian et al.~\cite{israelian04}).

Table~\ref{alfcen_params} lists the relevant physical properties of $\alpha$\,Cen~A and B,
and the astrometric and orbital parameters of the pair
are given in Table~\ref{alfcen_astrom_params}.
The position of the barycenter is computed from the Hipparcos astrometric solution of
components A and B using the masses of
Table~\ref{alfcen_astrom_params}. This gives a perfect consistency
with the Hipparcos data, but with an accuracy limited by the poor
astrometric solution of the B component. Another solution would
have been to use the astrometric solution of component A and the
 orbital elements of the system, whose uncertainty is hard to evaluate. The difference between the two
 approaches is about 0.02 arcsec, so it can be used as a good
 estimate of the uncertainty of the astrometric position of the
 barycenter in the ICRS frame at epoch J~1991.25.
 
\begin{table}
\caption[]{Properties of $\alpha$~Cen A and B.}
\label{alfcen_params}
\begin{tabular}{lcc}
\hline
\hline
& $\alpha$\,Cen\,A & $\alpha$\,Cen\,B \\
\hline
Other names & \object{HD 128620} & \object{HD 128621}  \\
& \object{HIP 71683} & \object{HIP 71681}  \\
\hline
$m_\mathrm{V}$ & -0.01 & 1.33\\
$m_\mathrm{J}$ & -1.16 & -0.01\\
$m_\mathrm{H}$ & -1.39 & -0.49\\
$m_\mathrm{K}$ & -1.50 & -0.60\\
Spectral Type & G2V & K1V\\
$\rm T_{\mathrm{eff}}$ (K)$^{\mathrm{a}}$ & $5790 \pm 30$ & $5260 \pm 50$ \\
$\log g$$^{\mathrm{a}}$  & $4.32 \pm 0.05$ & $4.51 \pm 0.08$ \\
$\left[ {\rm Fe/H} \right]$$^{\mathrm{a}}$  & $0.20 \pm 0.02$ & $0.23 \pm 0.03$ \\
\hline
\end{tabular}
\begin{list}{}{}
\item[$^{\mathrm{a}}$] From Morel et al.~(\cite{morel00}) and references therein.
\end{list}
\end{table}

\begin{table}
\caption[]{Astrometric parameters of $\alpha$\,Cen\,A-B.}
\label{alfcen_astrom_params}
\begin{tabular}{lcc}
\hline
\hline
Barycenter RA (epoch 1991.25, ICRS)$^{\mathrm{a}}$ & 14:39:40.216 \\
Barycenter Dec (epoch 1991.25, ICRS)$^{\mathrm{a}}$ & -60:50:13.58 \\
Proper motion RA (mas) & -3642.95 \\
Proper motion Dec (mas) & 694.75 \\
Radial velocity (km/s) & -20 \\
Galactic long. ($^\circ$) & 315.73 \\
Galactic lat. ($^\circ$) & -0.68 \\
Parallax (mas)$^{\mathrm{b}}$ & $747.1 \pm 1.2$ \\
Semi-major axis $a$ (")$^{\mathrm{c}}$ & $17.59$ \\
Period $P$ (yr)$^{\mathrm{c}}$ & $79.9$ \\
Excentricity $e$$^{\mathrm{c}}$ & $0.519$ \\
Inclination $i$ ($^\circ$)$^{\mathrm{c}}$ & $79.23$ \\
Long. of ascending node $\Omega$ ($^\circ$)$^{\mathrm{c}}$ & 204.82 \\
Longitude of pericenter $\omega$ ($^\circ$)$^{\mathrm{c}}$ & 231.80 \\
Reference epoch$^{\mathrm{c}}$ & 1955.59 \\
$M_A$ ($M_\odot$)$^{\mathrm{d}}$ & 1.10 \\
$M_B$ ($M_\odot$)$^{\mathrm{d}}$ & 0.91 \\
\hline
\end{tabular}
\begin{list}{}{}
\item[$^{\mathrm{a}}$] From Hipparcos (ESA~\cite{esa97}).
\item[$^{\mathrm{b}}$] Parallax from S\"oderhjelm~(\cite{soderhjelm99}).
\item[$^{\mathrm{c}}$] Orbital elements from Pourbaix~(\cite{pourbaix00}).
\item[$^{\mathrm{d}}$] Masses from Th\'evenin et al.~(\cite{thevenin02}).
\end{list}
\end{table}

\subsection{Orbital stability}

Presently, at least 15 examples of extrasolar planets are known to orbit binary star members:
16\,Cyg~B, $\upsilon$\,And, $\tau$\,Boo, etc... (Eggenberger, Udry \& Mayor~\cite{eggenberger03};
Eggenberger et al.~\cite{eggenberger04}; Mugrauer et al.~\cite{mugrauer05}).
The 40\,yrs period binary $\gamma$\,Cep is also very likely the host of a 1.3\,$M_J$
planet on a 1.8\,AU orbit (Cochran et al.~\cite{cochran02}).
Wiegert \& Holman~(\cite{wiegert97}) have identified how stable orbits
can be found within 2" of  $\alpha$\,Cen\,B (interior planets)
or at distances of up to 50" (exterior planets, orbiting the pair).
As a further incentive, it has been demonstrated that
the Kozai resonance (Holman et al.~\cite{holman97}; Innanen et al.~\cite{innanen97})
can prevent the ejection of a binary star companion
through chaotic variations in the excentricity of its orbit. This mechanism
is invoked by Mazeh et al.~(\cite{mazeh97}) to explain the presence of the planet
around 16\,Cyg\,B. High relative inclinations favor this mechanism, and
it can also be observed in the Solar system through the secular
perturbations introduced by Jupiter on asteroids (Kozai~\cite{kozai62}).
Such a dynamical, chaotic
behavior could stabilize the orbit of a BD around $\alpha$\,Cen\,B
beyond the maximum angular separation found by Wiegert \& Holman~(\cite{wiegert97}).
Recently, a hot Jupiter was detected around the primary star of the triple
system \object{HD 188753} (Konacki et al.~\cite{konacki05}).
The semi-major axis of the primary-secondary orbit is $a = 12.3$\,AU,
only half of $\alpha$\,Cen~A-B ($a = 23.7$\,AU).
Moreover, any angular separation can exist for a companion in
orbit around the $\alpha$\,Cen pair. This is such a favorable target for
deep imaging of its environment that it stands out as an
important step in testing the results of these numerical simulations.

\section{NACO adaptive optics imaging}
\label{obs}

\subsection{Observations}

We have chosen to adapt our observation technique depending on the angular distance to the star.
Very close to the two stars, within a radius of about 20", adaptive optics (subsequently AO) imaging allows us
to reach the highest sensitivity thanks to the concentration of the companion light within the Airy disk.
The contrast between the companion and the diffused light background is much more favorable
than for atmosphere-limited imaging.
At distances of more than 20", the diffused light is less of a problem, and
classical (non-AO) imaging is the best solution.
Moreover, the degradation of the AO correction quality
at such large distances from the star would not bring a
significant improvement in the sensitivity.
We will present our wide-field imaging observations
of the environment of $\alpha$\,Cen in a forthcoming paper.

We thus observed the environment of $\alpha$\,Cen\,B using the
Nasmyth Adaptive Optics System (NAOS, Rousset et al.~\cite{rousset00};
Rousset et al.~\cite{rousset03}) of the Very Large Telescope (VLT),
coupled to the CONICA infrared camera (Lenzen et al.~\cite{lenzen98}).
The combination of these two devices is abbreviated as NACO.
NAOS is equipped with a tip-tilt mirror and a deformable mirror controlled by
185 actuators, as well as two wavefront sensors: one for visible light and one for the infrared
domain. For our observations, we exclusively used the visible light wavefront
sensor. The detector is a $1024 \times 1024$ pixels ALADDIN InSb array.
As its name indicates, NACO is installed at the Nasmyth focus of the
Unit Telescope~4 (Yepun), the easternmost of the four 8\,m telescopes of the VLT.
Our observations were obtained shortly after the recoating of the primary
mirror, which was executed in October 2003, in order to benefit from the best possible
uniformity in reflectivity.
This excellent state of the primary mirror coating allowed us to minimize the
PSF light leaks, and consequently to obtain the best sensitivity.
The NACO instrument offers two coronagraphic modes, based on a classical Lyot
coronagraph or an innovative four-quadrant phase mask (Rouan et al.~\cite{rouan00}), but
due to the extreme brightness of $\alpha$\,Cen, the rejection level was
insufficient for preventing the saturation of the detector.
As a consequence, we preferred to use the direct
imaging mode and keep the two stars outside the detector.
This was achieved simply by offsetting the NACO field of view.

The first series of observations were obtained between February 18 and
April 10, 2004. We obtained repeated short exposures of four fields arranged in
a cross around B and (accidentally) one field East of A using the S13 mode of CONICA
and $J$$H$$K_s$ broadband filters.
The pixel scale in this mode is $13.26 \pm 0.03$\,mas/pix (Masciadri et al.~\cite{masciadri03}),
giving a field of view of 13.6"$\times$13.6".
This small scale results in an excellent sampling of the PSF, with $\simeq 5$\,pix/PSF,
an important advantage when distinguishing the point-like sources
from the speckle cloud based on their dimension and shape.
For all fields, the AO reference star was $\alpha$\,Cen\,B.

\begin{figure}[t]
\centering
\includegraphics[bb=0 0 360 144, width=8.7cm]{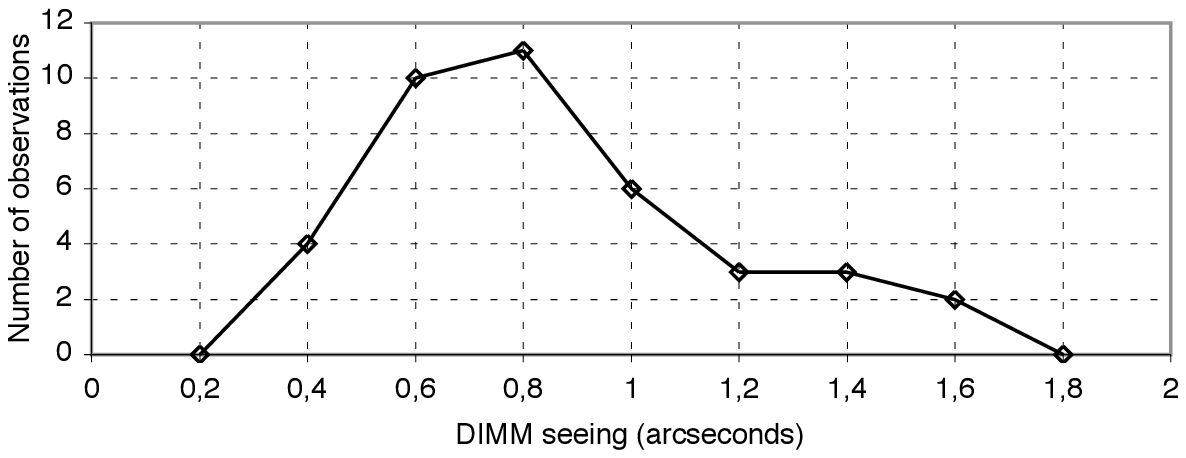}
\caption{Histogram of the number of NACO images as a function of DIMM seeing
(in visible light, with seeing bins of 0.2").}
\label{seeing_naco}
\end{figure}

\begin{figure}[t]
\centering
\includegraphics{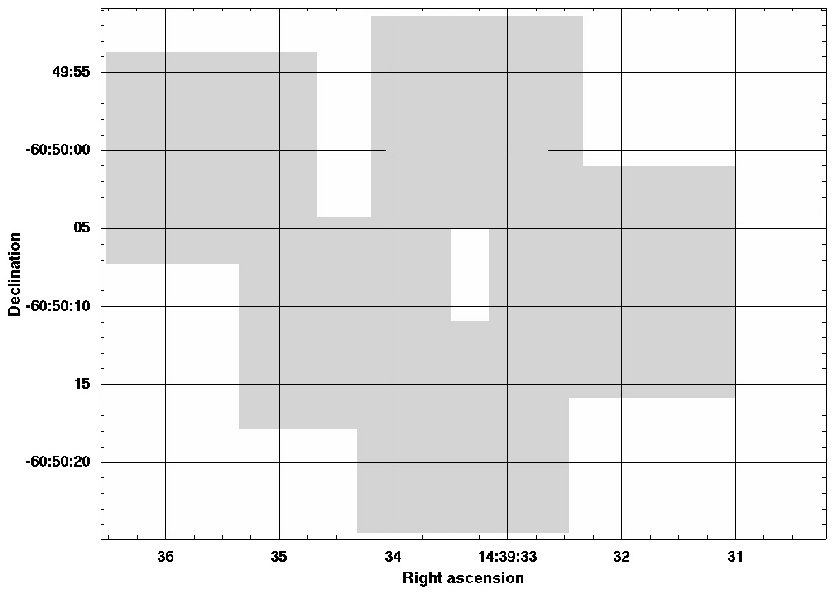}
\includegraphics{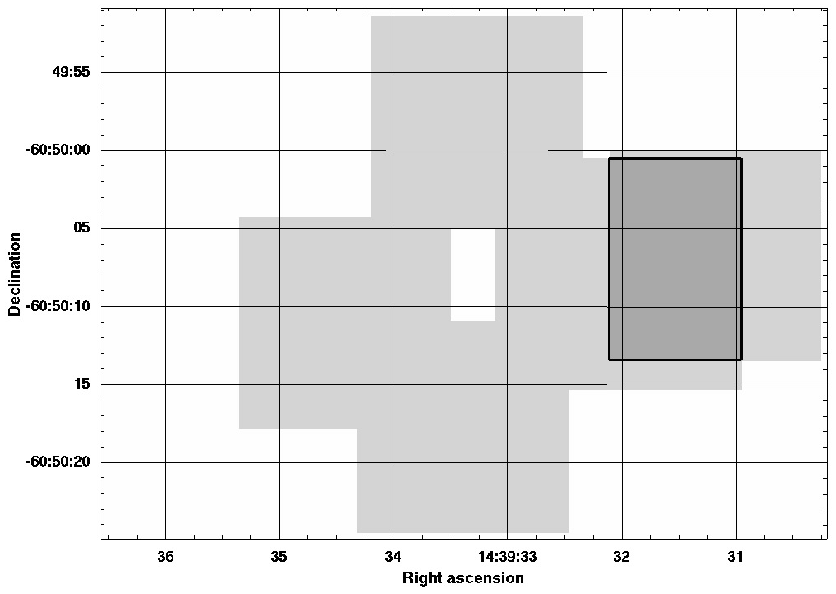}
\includegraphics{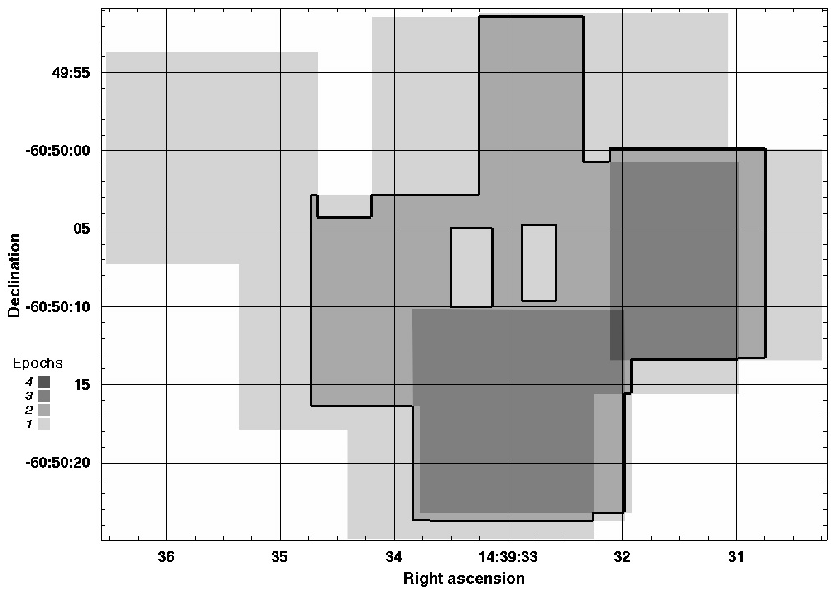}
\caption{Number of NACO epochs available for the field around $\alpha$\,Cen\,B
in the $J$ (top), $H$ (middle), and $K_s$ (bottom) bands.
The grey scale (bottom left) indicates how many epochs were obtained at each position.
The thick black line encompasses the domain that was observed at least on two epochs.}
\label{epochs_JHK}
\end{figure}

We repeated the same observations one year later in order to identify the proper-motion
companions, using the $K_s$ filter only because all the sources identified in the $J$ and $H$
bands were also detected in $K_s$. One image of the southern field was obtained
in July 2004, but due to operational constraints, the remaining observations were
conducted in February-March 2005.
All observations were obtained using the Fowler-sampling, high-sensitivity mode of CONICA.
The individual exposures were limited to 5.0\,s in the $J$ band and 3.5\,s in the $H$ and $K_s$ bands.
The complete observation log is presented in Tables~\ref{naco_log1}, \ref{naco_log2}, and \ref{naco_log3},
together with the seeing\footnote{available at http://archive.eso.org/asm/ambient-server}
observed in the visible by the DIMM (Sarazin \& Roddier~\cite{sarazin90}; Martin et al.~\cite{martin00}).
Our NACO images were obtained in general under good seeing conditions,
many of them at or below the 0.7" level in the visible (Fig.~\ref{seeing_naco}).
This is a clear advantage in detecting very faint sources as the coherent
energy (encircled in the core of the diffraction-limited PSF) can then rise up to 70\% or more.
In each field, the total integration time for each epoch varies between 10 and 20\,minutes.
Depending on the location around $\alpha$\,Cen, up to four epochs are available.
The resulting total coverage in the $J$, $H$ and $K_s$ bands is presented in
Fig.\,\ref{epochs_JHK}. The $H$ and $J$ band data cover a comparatively smaller area
and only one epoch was obtained in the $J$ band due to the relatively lower
sensitivity compared to $H$ and $K_s$.
When only one epoch was obtained, the search for comoving companions is not
possible (see also Sect.~\ref{companions}).

\begin{table}\caption{Log of the first series of NACO images.}
\label{naco_log1}
\begin{tabular}{lcclccc}
\hline
Field & Date & UT$^a$ & $\lambda^b$ & $\delta t$(s)$^c$ & $\theta$(")$^d$ & AM$^e$ \\
\hline
West B & 2004-02-18 & 7:09 & $J$ & 360 & 0.48 & 1.38 \\
West B & 2004-02-18 & 7:15 & $H$ & 392 & 0.49 & 1.37 \\
West B & 2004-02-18 & 7:21 & $K_s$ & 322 & 0.49 & 1.35 \\
South B & 2004-02-18 & 8:10 & $J$ & 480 & 0.51 & 1.28 \\
South B & 2004-02-18 & 8:15 & $H$ & 525 & 0.56 & 1.28 \\
South B & 2004-02-18 & 8:20 & $K_s$ & 518 & 0.51 & 1.27 \\
North B & 2004-02-18 & 9:03 & $J$ & 320 & 0.52 & 1.24 \\
North B & 2004-02-18 & 9:08 & $H$ & 500 & 0.45 & 1.24 \\
North B & 2004-02-18 & 9:14 & $K_s$ & 500 & 0.54 & 1.24 \\
East B & 2004-02-19 & 8:33 & $J$ & 480 & 1.07 & 1.26 \\
East B & 2004-02-19 & 8:54 & $H$ & 525 & 0.92 & 1.25 \\
East B & 2004-02-19 & 8:59 & $K_s$ & 518 & 0.75 & 1.24 \\
West B & 2004-02-20 & 6:47 & $J$ & 360 & 0.72 & 1.41 \\
West B & 2004-02-20 & 6:52 & $H$ & 392 & 0.68 & 1.40 \\
West B & 2004-02-20 & 6:59 & $K_s$ & 322 & 0.86 & 1.38 \\
South B & 2004-02-20 & 7:39 & $J$ & 480 & 1.03 & 1.31 \\
South B & 2004-02-20 & 7:44 & $H$ & 525 & 1.03 & 1.30 \\
South B & 2004-02-20 & 7:50 & $K_s$ & 518 & 0.85 & 1.30 \\
North B & 2004-02-20 & 8:38 & $J$ & 480 & 0.74 & 1.25 \\
North B & 2004-02-20 & 8:43 & $H$ & 750 & 0.71 & 1.25 \\
North B & 2004-02-20 & 8:49 & $K_s$ & 740 & 0.71 & 1.25 \\
East B & 2004-02-26 & 7:58 & $J$ & 160 & 1.62 & 1.26 \\
East B & 2004-02-26 & 8:03 & $H$ & 175 & 1.28 & 1.26 \\
East B & 2004-02-26 & 8:08 & $K_s$ & 175 & 1.53 & 1.26 \\
East A & 2004-03-12 & 7:59 & $H$ & 175 & 1.37 & 1.24 \\
East A & 2004-03-12 & 8:02 & $J$ & 320 & 1.35 & 1.24 \\
East A & 2004-03-12 & 8:04 & $K_s$ & 175 & 1.35 & 1.24 \\
East B & 2004-04-10 & 5:14 & $J$ & 480 & 0.61 & 1.26 \\
East B & 2004-04-10 & 5:23 & $H$ & 525 & 0.58 & 1.25 \\
East B & 2004-04-10 & 5:29 & $K_s$ & 518 & 0.57 & 1.25 \\
\hline
\end{tabular}
\begin{list}{}{}
\item[$^{\mathrm{a}}$] Average time of the observation sequence.
\item[$^{\mathrm{b}}$] Selected broadband filter.
\item[$^{\mathrm{c}}$] Total exposure time.
\item[$^{\mathrm{d}}$] DIMM seeing measured in the visible ($\lambda=0.5\,\mu$m).
\item[$^{\mathrm{e}}$] Airmass.
\end{list}
\end{table}


\begin{table}\caption{Second series of NACO images.}
\label{naco_log2}
\begin{tabular}{lcclccc}
\hline
Field & Date & UT & $\lambda$ & $\delta t$(s) & $\theta$(") & AM \\
\hline
South B & 2004-07-25 & 1:29 & $K_s$ & 1008 & 0.99 & 1.37 \\
West B & 2005-02-07 & 7:24 & $K_s$ & 535.5 & 1.10 & 1.44 \\
West B & 2005-02-09 & 8:22 & $K_s$ & 535.5 & 1.05 & 1.31 \\
South B & 2005-02-09 & 8:50 & $K_s$ & 756 & 1.27 & 1.27 \\
West B & 2005-02-09 & 9:46 & $K_s$ & 535.5 & 0.95 & 1.24 \\
East B & 2005-03-28 & 5:21 & $K_s$ & 567 & 0.63 & 1.30 \\
South B & 2005-03-29 & 4:04 & $K_s$ & 756 & 0.87 & 1.45 \\
North B & 2005-03-29 & 4:31 & $K_s$ & 216 & 0.76 & 1.38 \\
\hline
\end{tabular}
\end{table}

\begin{table}\caption{Third series of NACO images.}
\label{naco_log3}
\begin{tabular}{lcclccc}
\hline
Field & Date & UT & $\lambda$ & $\delta t$(s) & $\theta$(") & AM \\
\hline
West B & 2005-07-13 & 23:30 & $K_s$ & 787.5 & 0.75 & 1.24 \\
West B & 2005-07-13 & 23:53 & $H$ & 756 & 0.82 & 1.24 \\
\hline
\end{tabular}
\end{table}

\subsection{Data processing}

The processing was achieved using the IRAF package (v.2.12).
The images obtained on each night were dark-subtracted and flat-fielded with
standard infrared astronomical techniques. No sky subtraction was done, as the
inhomogeneous diffused light from $\alpha$\,Cen largely dominates
the sky background level, even in the $K_s$ band (see Sect.~\ref{diffused}).
We interpolated the bad pixels and mosaicked the dither pattern using
the bright sources visible in each field as references. The observed drift rate remains
below 30\,mas.hr$^{-1}$ for all fields (less than 2.5 pixels, or half the FWHM of the PSF).
This is a remarkable performance for such a large and massive instrument
and telescope configuration. For the northern field in the $J$ band, no source
was detected in each individual frame, so the registration was achieved using the
offsets measured on the $H$ and $K_s$ band images. This procedure is
justified by the fact that the observed relative drifts were identical in the three bands.
In any case, no source was detected in the northern field in the combined $J$
band image.
Bright artefacts (``ghosts") are present in the images of the fields located to the west
of both stars. They are probably caused by reflections and interferences in the semi-reflective
beam splitters used to separate the visible light (used for wavefront sensing)
from the infrared.

\subsection{Astrometric calibration}
\label{astrometry}

The astrometric referencing of narrow-field NACO images of a
fast-moving source such as $\alpha$\,Cen ($\approx 4"/$yr) is not a straightforward task.
It is made all the more difficult as all direct images of the pair are heavily saturated.
As a consequence, there are no astrometric reference stars sufficiently
close to the pair to attach the NACO images to a solid astrometric reference.

\begin{figure*}[t]
\centering
\includegraphics{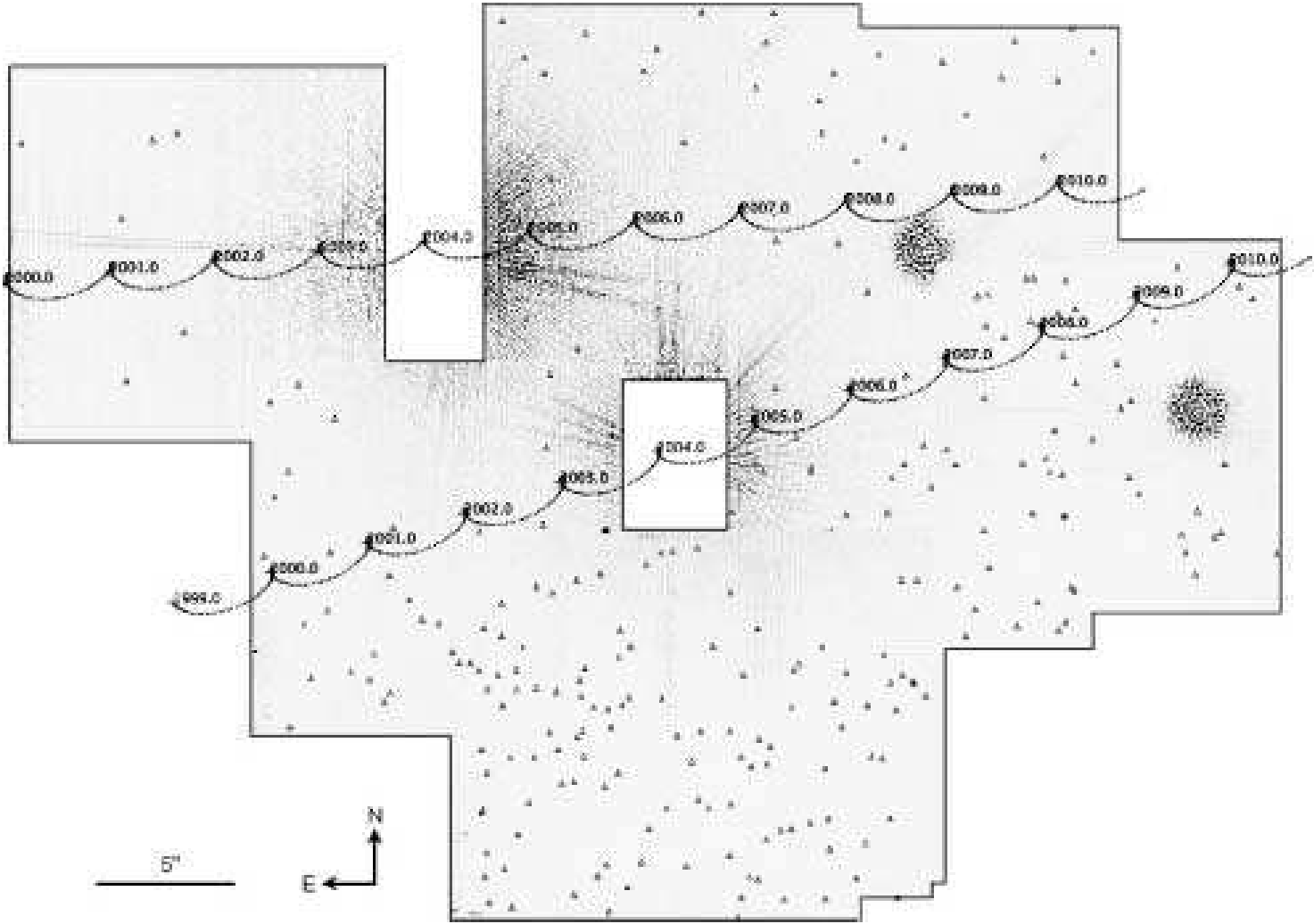}
\caption{Mosaic of the observed NACO fields of the environment of $\alpha$\,Cen
in the $K_s$ band.
This mosaic is a composite of the filtered versions of the original images,
using the ring median filtering described in the text.
The apparent ICRS positions on the sky of $\alpha$\,Cen A and B,
plotted for the 1999-2010 period, include their proper motion, orbital motion, and
parallactic apparent displacement. The detected sources are represented as open
triangles. The two circular features located to the west of the image are instrumental artefacts.}
\label{alfcenpm}
\end{figure*}

We thus based our absolute astrometric calibration on the computed
positions of the $\alpha$\,Cen\,B star using the initial Hipparcos
position at J1991.25 to determine the ICRS astrometric position of the
barycenter. Then the positions of $\alpha$\,Cen\,B are
computed at any other epoch with the orbital elements
from Pourbaix~(\cite{pourbaix00}) and proper motion measurements
from the {\it Hipparcos} satellite (ESA~\cite{esa97}).
These positions take into
account the combined effect of the parallactic apparent
displacement, proper motion, and orbital motion of the pair. It
should be noted that, even though the resulting apparent
displacement is particularly complicated, the accuracy of the
available astrometric elements is such that they do not limit the
astrometric calibration of our images. As an illustration of the
complexity of the apparent motion of $\alpha$\,Cen A and B,
Fig.~\ref{alfcenpm} shows the ICRS positions of the two stars on
the sky for the period 1999-2010.

To transfer the reference coordinates of $\alpha$\,Cen\,B to the detected
background sources, we used as an intermediate step the HST-ACS coronagraphic
images publicly available from the ST-ECF archive facility. These images present
the advantage of simultaneously showing an attenuated image of the occulted
$\alpha$\,Cen\,B, as well as several background sources that are detectable
on our NACO images. We chose the brightest of these sources, which is also the nearest
to $\alpha$\,Cen\,B, numbered 167 in our catalogue (Table~\ref{naco_objects1}).
This ACS image was obtained on 15 June 2004, for which date we computed the ICRS
position of B to be\\
$\alpha(B) = $ 14:39:32.953\ \ $\delta(B) = $ -60:50:08.65,\\
thus giving the following coordinates for star \#167:\\
$\alpha(\#167) = $ 14:39:33.567\ \ $\delta(\#167) =$ -60:50:11.57.\\
Knowing the absolute sky coordinates of each star, a "world
coordinate system" (wcs) was defined for each image based on the measured
position, the pixel scale of the camera, and the orientation of the field of view
($y$ axis aligned along the north-south direction). The relative astrometric accuracy
over the HST-ACS field is better than 5\,mas, therefore introducing a negligible
uncertainty in the coordinate transfer.

We would like to stress here that the wcs used for all
our images is linked to the computed position of $\alpha$\,Cen\,B on 15 June 2004.
Any modification of the computed astrometric position
of $B$ for this date can be transferred to the source catalogue using a simple
translation. Given the small size of the field, we expect 
an absolute astrometric accuracy better than $\pm 0.10$"
from this very simple astrometric reduction.
However, the relative position accuracy of the different sources within the
same NACO field is much better, with an estimated $\pm 0.03$" (2 pixels).
The orientation of true north of the S13 camera of NACO was found to be extremely
stable and accurate by Chauvin et al.~(\cite{chauvin05}), with an undetectable deviation of
less than 0.1$^\circ$ from the true north-south direction over a period
of more than one year (Nov.~2002-Mar.~2004). The uncertainty of the scale is
estimated to be less than 0.2\%, giving at most one pixel over a
field of 30 arcsec.

\begin{figure}[t]
\centering
\includegraphics{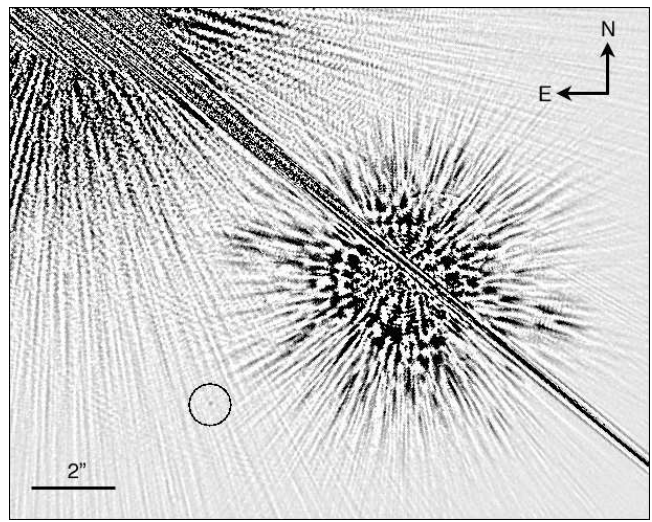}
\caption{Extract of an HST-ACS coronagraphic image of $\alpha$\,Cen\,B showing the
bright background star \#167 (circled) that was used to transfer the reference astrometric
coordinates of $\alpha$\,Cen\,B to the NACO images. The saturated source in the upper left
corner is $\alpha$\,Cen~A.}
\label{acs_b}
\end{figure}

\subsection{Diffused light}
\label{diffused}

The main limitation to the sensitivity of imaging close to bright sources is
caused by diffused light. It is mostly created inside the telescope
and the instrument by imperfect optics and baffling.
For the preparation of adaptive optics observations of bright sources,
it is important to know the properties of the diffused light to
prevent saturation of the detector.

\begin{figure}[t]
\centering
\includegraphics[bb=0 0 360 288, width=8.7cm]{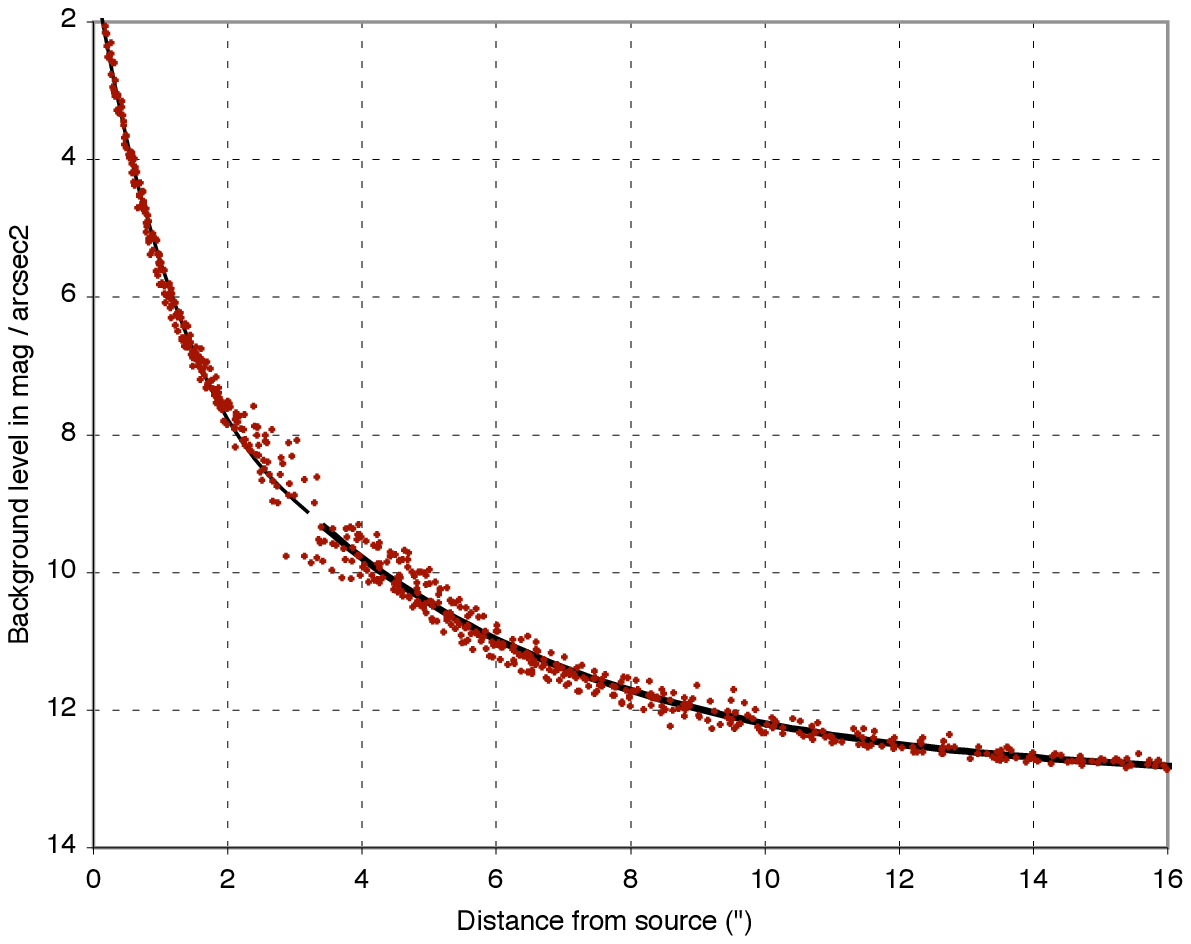}
\caption{Model fitting of the NACO diffuse background in the $K_s$ band.
The dots represent the individual measurements on the image, scaled to a zero-magnitude star,
and the superimposed curves are the fitted models near to (thin curve) and far from (thick curve) the
central star.}
\label{diffused_light_Ks}
\end{figure}

To study its profile in our images,
we considered the field located south of $\alpha$\,Cen\,B,
thereby avoiding the contamination by the light from $\alpha$\,Cen\,A,
which is particularly strong in the northern and eastern fields.
A difficulty in measuring the diffused light is that a number of artefacts create
local biases. For instance, the large spikes produced by the secondary spider and the
ghost reflections visible in the western images should not be included in the background
estimation. We thus sampled the background level manually to avoid these
artefacts.
The result was a series of $\approx 500$ samples $N(\theta)$ in each band,
with $\theta$ the angular distance from $\alpha$\,Cen\,B and $N$ the camera counts (in ADUs).
These measurements were then converted to magnitudes per squared arcsec
taking into account the exposure time ($\delta t = 5.0$\,s in $J$, 3.5\,s in $H$
and $K_s$), the pixel size ($\delta\theta=0.01326$"), and the photometric
zero point for the night (ZP$_J$=23.95, ZP$_H$=23.85, ZP$_{Ks}$=22.95).
In order to obtain a calibrated model that can be applied to other
sources, we normalized the resulting magnitudes to a zero magnitude source using the
apparent magnitudes $m_\lambda$(B) of $\alpha$\,Cen\,B listed in Table~\ref{alfcen_params}.
The expression of the measured sky-backgound contrast (in mag.arcsec$^{-2}$) is therefore:
\begin{equation}
\Delta m_\lambda(\theta) = -2.5 \log \left[\frac{N(\theta)}{\delta t\ \delta\theta^2}\right] + {\rm ZP}_\lambda
- m_\lambda \left({\rm B} \right).
\end{equation}
We subsequently computed a least-square fit to our data using an exponential model of the form:
\begin{equation}
\Delta m_\lambda(\theta) = a\,\exp( b \, \theta ) + c.
\end{equation}
As shown in Fig.~\ref{diffused_light_Ks}, this type of model is a good match to the observed
distribution. The resulting best-fit values of the $(a,b,c)$ coefficients are listed for each band
in Table~\ref{naco_diffused}.
The profiles obtained in the three bands are very similar,
and they show a relative flux level with respect to the central source
of $\Delta m \approx 9$\,mag.arcsec$^{-2}$ at a distance of 3".
To extend our diffused background model closer to the central star, we took advantage
of the unsaturated acquisition images (left part of Fig.~\ref{diffused_light_Ks}).
It is interesting to note that the estimated brightness at a large distance from the star
tends asymptotically to $m_K \approx 13$\,mag/arcsec$^{2}$, which is close to
the typical Paranal sky brightness level in the $K$ band.

\begin{table}
\caption{Diffused light model parameters. We considered an exponential model
$\Delta m_\lambda = a\ e^{b \theta} + c$, where $\Delta m_\lambda$ is the surface magnitude
per arcsec$^2$ of the sky background at an angular distance $\theta$ of a
zero magnitude source, and $(a,b,c)$ are the adjusted model parameters. }
\label{naco_diffused}
\begin{tabular}{lcccc}
\hline
$\lambda$ & $J$ ($\theta \ge 3$") & $H$ ($\theta \ge 3$") & $K_s$ ($\theta \ge 3$") & $K_s$ ($\theta \le 3$") \\
\hline
$a$ & -7.281 & -7.969 & -8.035 & -9.068 \\
$b$  & -0.163 & -0.194 & -0.228 & -0.638 \\
$c$  & 13.226 & 13.247 & 13.018 & 10.305 \\
\hline
\end{tabular}
\end{table}

\subsection{Source extraction and photometry}
\label{filtering}

The greatest difficulty in extracting sources from the diffused light of
$\alpha$\,Cen is to separate the background inhomogeneities from the
true point-like sources.
We first high-pass-filtered the combined images
using the ring median filter of IRAF (Secker~\cite{secker95}).
By adjusting the ring radius precisely to the radius of the PSF,
it is possible to isolate the smooth, low spatial frequency diffused light
and remove it from the image.
This filtering allows a much more robust identification
of the point-like sources.
With only 252 sources in total, a visual identification was found to be more efficient
than an automated detection algorithm.
We used the $K_s$ band images for this identification,
as all sources detected in $J$ and $H$ were also detected in this band.

The difficulty with automated source identification is to adapt the sensitivity
to the rapidly changing background level depending on the distance to the
star. The identification of the sources was thus achieved using the blinking of the
ring median-filtered versions of the combined NACO images obtained in the $J$, $H$,
and $K_s$ bands. The availability of images obtained through several filters is
a big advantage, as the fixed speckle cloud scales with the observation
wavelength.

We derived aperture photometry for the detected sources using IRAF.
We also attempted the PSF-fitting technique, but due to the large perturbations
of the PSF shape by high spatial-frequency speckles close to the two stars,
the result of the star subtraction was not satisfactory. We chose tight apertures
of 24, 12, and 10 pixels in diameter (0.32", 0.16", and 0.13"), respectively for the $J$, $H$,
and $K_s$ bands, in order to reduce our sensitivity to the background fluctuations.
By using such small apertures, we became more sensitive to the quality of the AO correction;
but thanks to the brightness of the source and the generally good seeing,
the Strehl ratio was relatively stable over our observations.
The background level itself was estimated from the median flux of
a ring of 50, 30, and 20 pixels in diameter (respectively for $J$, $H$, and $K_s$)
and 10 pixels in thickness (in all cases).

The computation of the aperture correction
was achieved on one of the brightest sources of our catalogue in the western field
(star \#27 in Table~\ref{naco_objects1}).
This source is located far enough (12") from $\alpha$\,Cen\,B
so that the local background can be considered as flat.
The resulting aperture correction of $\Delta m=1.0 \pm 0.2$, consistent in all three bands,
was applied to the derived magnitudes.
The photometric zero points were taken
from ESO's routine instrument monitoring program. This is justified by the
fact that the nights during which our observations were obtained were all of photometric quality.
Considering the inhomogeneity of the background over which the photometry is
measured, the contribution from the uncertainty to the photometric zero points is negligible.
The airmass corrections were neglected: using the 2MASS values determined by
Nikolaev et al.~(\cite{nikolaev00}; $A_J=0.092$, $A_H=0.031$,
$A_K=0.065$\,mag/airmass, relative to unity airmass),
they are always smaller than 0.05 magnitudes in all three bands.
In order to account for the Strehl ratio fluctuations, background inhomogeneity, and
aperture correction uncertainty, a conservative systematic error was added to the statistical error
bars listed in Table~\ref{naco_objects1}: $\pm 0.8$\,mag on
the $J$ band magnitudes (due to the relatively stronger diffused background)
and $\pm 0.5$\,mag on the $H$ and $K_s$ magnitudes. 

\subsection{Sensitivity}

\begin{figure}[t]
\centering
\includegraphics[bb=0 0 360 144, width=8.7cm]{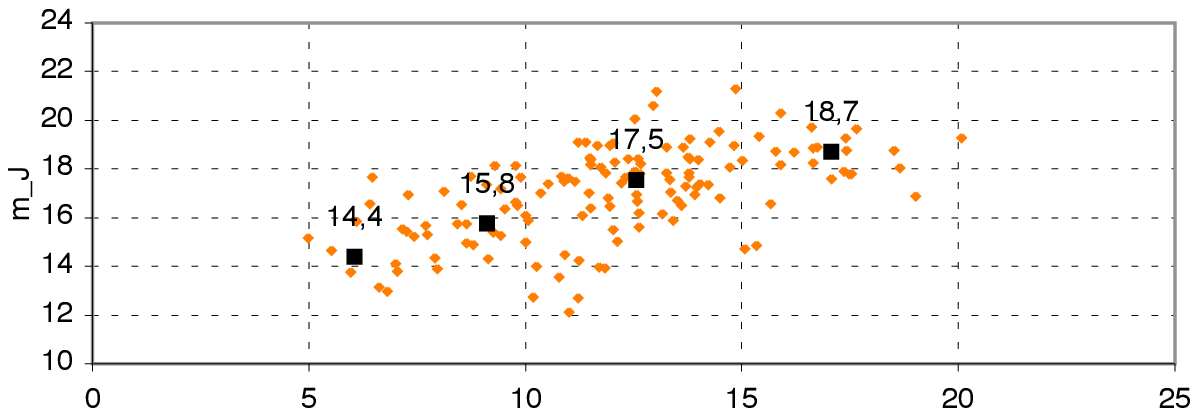}
\includegraphics[bb=0 0 360 144, width=8.7cm]{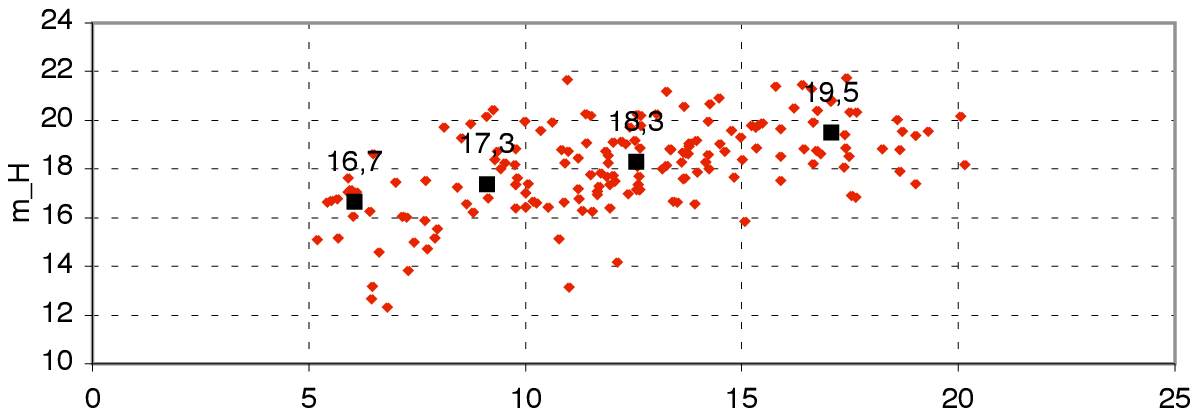}
\includegraphics[bb=0 0 360 144, width=8.7cm]{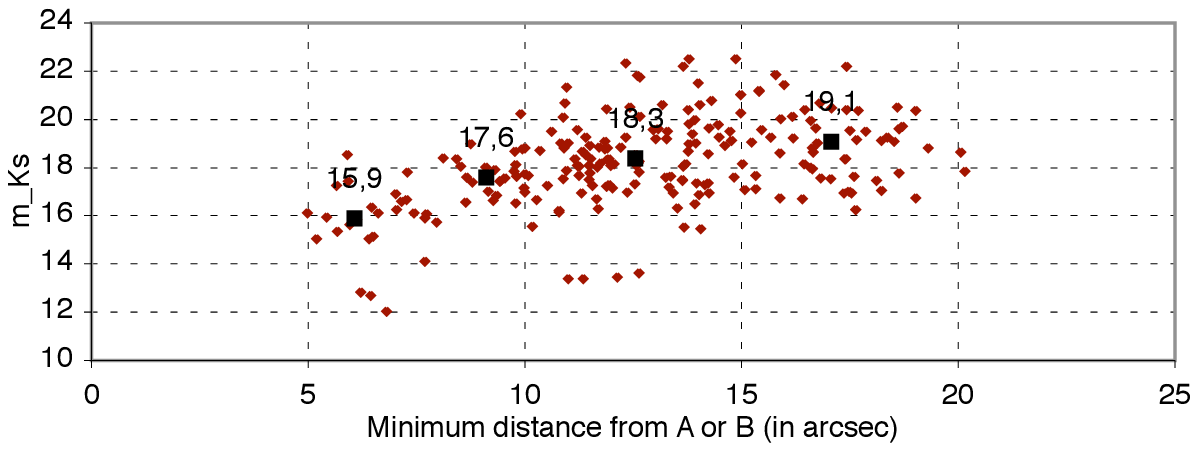}
\caption{Apparent magnitudes of the detected objects, as a function of their
minimum angular separation with $\alpha$\,Cen~A or B. The solid squares
correspond to the median magnitude of the objects detected in the
$5-7$", $7-10$", $10-15$", and $15-20$" domains, respectively. The corresponding
magnitude is indicated in each case.}
\label{mag_objects}
\end{figure}

The definition of the sensitivity of our search for companions around a binary star like
$\alpha$\,Cen is more difficult than for a single star.
The presence of the combined diffused light from $\alpha$\,Cen~A
and B in the NACO fields complicates the estimation of limiting magnitudes,
as they become dependent on the position relative to the
two bright stars.

We thus preferred to take advantage of the significant number of detected
sources to derive {\it a posteriori} statistical properties and estimate the true
sensitivity of our imaging survey. As shown in Fig.~\ref{mag_objects},
the magnitudes of the faintest sources at angular distances larger than 10" from A and B
are $\approx 20$ in the $J$ band and $\approx 21$ in the $H$ and $K_s$ bands.
Note that several very faint and/or close-in sources, while they were clearly
detected in the images, could not be measured by aperture photometry.
To define our practical limiting magnitude,
we chose to consider the median magnitude of the detected sources.
This definition has the advantage of giving an empirical, statistically meaningful
definition of the sensitivity, which can be expressed as a function of the distance to the
two bright stars by computing the median within angular distance bins.
Figure~\ref{mag_objects} shows the median of the detected object magnitudes in the $J$, $H$
and $K_s$ bands for four angular distance bins: $6-7$", $7-10$", $10-15$", and $\ge 15$" as solid
squares. The retained angular distance is the minimum of the source distances to
A and B so as to account for the ``saddle" shape of the diffused light from the two stars.
In all bands, the limiting magnitudes at large angular distances are in the $18-20$ range.
They decrease to $m_J = 14.4$, $m_H=16.7$, and $m_{Ks}=15.9$ at 6".

We can compare these sensitivities with previous AO studies of the environment
of bright stars.
Using the same NACO instrument,
Chauvin et al.~(\cite{chauvin05}) obtained a depth of $m_{Ks}=20$ around HIP\,6856
($m_{Ks}=6.8$) using an exposure of $10 \times 15$\,s.
With our typical exposures of $150 \times 3.5$\,s, we are affected by additional
readout noise, but the longer total exposure time compensates for this loss.
On the bright single star Vega ($m \approx 0$ in all bands),
Metchev et al.~(\cite{metchev03}) used the PALAO system installed at the 5\,m Hale
reflector at Mount Palomar in the $J$, $H$ and $K_s$ bands.
With limiting magnitudes of $m_H \approx 18$ at 20" and $\approx 16$ at 10",
their study is slightly less sensitive than ours, but this can be explained by the smaller aperture.
Macintosh et al.~(\cite{macintosh03}) observed the same star using
the Keck AO system and reach a deeper $m_{Ks} \approx 20.5$ at 20", $18.5$ at 10",
and $17$ at 7", using a $90 \times 15$\,s exposure. These figures are comparable
to our results, although Vega is fainter by about 0.5\,mag than $\alpha$\,Cen\,B in
the infrared.
From these comparisons, it appears that
NACO is a well-suited instrument for studying the environment of bright stars,
as its diffused light signature is relatively low (see also Sect.~\ref{diffused}).
In addition, the structure of the fixed-pattern speckle halo
created by the monolithic primary mirror of the VLT-UT4 telescope
appears smoother than with the Keck telescope's segmented primary mirror,
thus making the identification of close companions easier.

\section{HST archive data \label{hst_archive}}

As a complement to our NACO images, 
we searched the ESO/ST-ECF archive for images of $\alpha$\,Cen.
We subsequently analyzed the available data, that were obtained
using three HST instruments: ACS, NICMOS and WFPC2. In this Section,
we present briefly our results.

\subsection{ACS \label{hstacs}}

A series of images was obtained centered on $\alpha$\,Cen~A star
in September 2003 using the Advanced Camera for Surveys (ACS) onboard the
Hubble Space Telescope, and these observations were repeated in January 2004
to check for the presence of proper motion companions.
The same repeated series of images were obtained
for $\alpha$\,Cen\,B behind the coronagraphic mask in June 2004 and August 2004.
In each case, eight images were recorded at eight wavelengths
between $\lambda = 754$ and 1024\,nm, with the FR914M broad ramp filter
wheel (bandwidth of 9\%).

Using the coronagraphic mode of this instrument,
the principle of the foreseen data analysis was to use the fact that the PSF of the
instrument changes homothetically with the wavelength to remove
most of the fixed-pattern speckle noise. As the position of the potential
companions does not depend on the wavelength, their signature can
be extracted more efficiently from the speckle noise than with a single image.
This method is a particular application of the spectral deconvolution technique 
developed by Sparks \& Ford~(\cite{sparks02}).
However, only one star of the $\alpha$\,Cen pair
at a time can be aligned with the coronagraphic spot. This results in a
considerable amount of diffused light from the other, non-masked star, which also
scales with the wavelength but with a different homothetic center.
The application of the spectral deconvolution method is also made
difficult by the slight undersampling of the PSF and by the availability
of narrow-band filters instead of the continuous spectral coverage
provided by a dispersive spectro-imaging instrument.

The pre-processing was achieved using the automated pipeline available
at the HST archive. The images were subsequently co-added and filtered
using the same procedure as the NACO images (Sect.~\ref{filtering}).
An extract of the $\alpha$\,Cen\,B centered co-added coronagraphic image
is presented in Fig.~\ref{acs_b}.
Over a total field comparable to our NACO images, the number of
objects detected in the ACS images is less than 10\% of the NACO catalogue,
corresponding to the brightest objects. We therefore limited our use of the ACS
images to the definition of an accurate astrometric coordinate
system (see Sect.~\ref{astrometry}).

\subsection{NICMOS}

The HST-NICMOS instrument (Thompson et al.~\cite{thompson98}) is
based on an infrared  HgCdTe 256x256 pixel array sensitive over the
0.8-2.5~$\mu$m range.
Two series of exposures were taken through four
filters on 19 October 1998 ($\alpha$\,Cen\,B)
and 22 October 1998 ($\alpha$\,Cen\,A). The star images were positioned on the detector
surface without a coronagraphic mask, producing heavy saturation within a
radius of 2-3\,arcsec around each star.
The absence of a second observation makes it impossible
to ascertain the comoving nature of potential companions. The complexity
of the HST-NICMOS PSF (Krist et al.~\cite{krist98}) limits
the sensitivity close to the star. Being too distant
in time, there is no overlap between our NACO fields and these HST-NICMOS
data. For these reasons, we decided not to include the NICMOS data in
the present study.

\subsection{WFPC2}

The Wide Field and Planetary Camera 2 (WFPC2) is a two-dimensional imaging
photometer that covers the spectral range between 115 to 1050\,nm. Several
accepted GTO and open time proposals, in particular by Ford et al. and Henry et al.
in HST Cycles 4 to 7 resulted in a large amount of collected data. A total of 11
images centered on $\alpha$\,Cen~A were obtained in 1995 over two
epochs (around May and August) in the F547M, F555W, F814W, and F850LP filters.
In 1997, another series of 10 images was recorded, this time through the F953N
and F1042M filters. The same sequences were also obtained with the WFPC2
field centered on $\alpha$\,Cen\,B. As for the NICMOS data, these observations
are too far in time from our NACO images, and there is almost no overlap between
the fields. Therefore, we preferred not to include them in this study.

\section{Catalogue of the detected sources}
\label{cat}

Table~\ref{naco_objects1} lists the positions and $J$$H$$K_s$ 
magnitudes of all the sources detected in the NACO images of $\alpha$\,Cen.
The right ascension $\alpha$ and declination
$\delta$ refer to the ICRS and are not corrected for
possible parallax. The epoch is J2004.5, corresponding to the mean
observation time for stars observed in successive frames. As
explained earlier, the typical positional uncertainty is not larger than 0.1\,arcsec.
The relatively high surface density of the detected objects can be explained
by the fact that $\alpha$\,Cen lies almost exactly in the Galactic plane
and in a direction close to the Galactic center. This catalogue fills part of a
long-standing ``hole" in sky atlases, due to the diffused light from $\alpha$\,Cen.

\section{Proper-motion companion search \label{companions}}

The very fast proper motion of $\alpha$\,Cen should allow its
comoving companions to be identified quickly.
However, this is also a drawback due to the particularly
dense star field around this binary star.
The identification of the companion is not a trivial task
because of the combination of the unknown orbital motion
of the putative companion with the large proper motion and parallactic
displacement of $\alpha$\,Cen. Considering that $\alpha$\,Cen moves an average
of approximately one NACO pixel per day, the best strategy would to observe the fields
repeatedly with a time separation of 2 to 3 weeks. Unfortunately, due to scheduling
constraints, our observations could not follow this scheme, and our first and second
epochs were separated by about 10 months. The second and third series were
separated by 5 months. Over these durations, the displacement of $\alpha$\,Cen was
considerable, resulting in a rather poor overlap of the different fields. Moreover, the
diffused light from the two stars resulted in a moving zone of decreased sensitivity
over part of the field.

In order to systematically search for statistically significant proper-motion
companions, we applied the following procedure:
\begin{enumerate}
\item We converted the intensity images into SNR images, using the local background noise
(mostly made of residual speckles). This allowed us to select only the sources that present
an SNR of more than 3 per pixel, compared to the local noise, and a PSF shape
(first {\it a priori}). For each epoch and each color, we therefore obtained a map of all
point-like sources above the local noise.
\item The sources that could be identified at the same position on the sky
at different epochs are background sources, so they were eliminated from our sample.
\item The second {\it a priori} knowledge that we can use is that any companion to
$\alpha$\,Cen will move on a Keplerian orbit. Therefore, its maximum orbital motion
rate $\gamma$ is set by the third Kepler law and can be expressed as a function
of the angular distance $\theta$ from the star:
 \begin{equation}
 \gamma = \sqrt{ \frac{G\,M_*}{\theta\,d^3} }
 \end{equation}
where $d$ is the distance of the star, $M_*$ its mass, and $G$ the universal gravitational constant.
A numerical application for $\alpha$\,Cen results in the following maximum
orbital motion:
 \begin{equation}
 \gamma = \frac{3.83}{\sqrt{\theta}}
 \end{equation}
where $\gamma$ is in arcsec/yr and $\theta$ in arcsec.
Between two measurement epochs, we can therefore define a "possible orbital motion
disk", centered on each identified point source, whose radius depends on the time lapse
between the two epochs. The intersection of these disks with the
list of identified sources in the following epoch allowed us
to significantly reduce the number of candidate companions to just a few.
\item Eventually, the careful examination of the time evolution of the positions
of the residual candidate companions allows the unphysical orbits to be rejected.
\end{enumerate}
From this selection process, we could not identify any comoving companion within
our overlapping regions.
For only one source we were able to obtain a significant detection (more than 4$\sigma$ per
pixel above the local noise) at epoch 2005.104, while no source was apparent
at epoch 2004.137. Its ICRS coordinates are $\alpha=$14:39:32.118 $\delta=$-60:50:08.60,
and its magnitude is estimated at $m_{Ks}=17.7 \pm 0.5$.
However, we could not identify any counterpart of this source within the ``orbital circle"
of angular radius $\gamma$, as defined in step 3 of our identification procedure.
This source could be a distant variable star or a faint solar system object.
We have not included it in the catalogue due to its unconfirmed nature.

\section{Discussion}
\label{discussion}

\begin{figure}[t]
\centering
\includegraphics[bb=0 0 360 288, width=8.7cm]{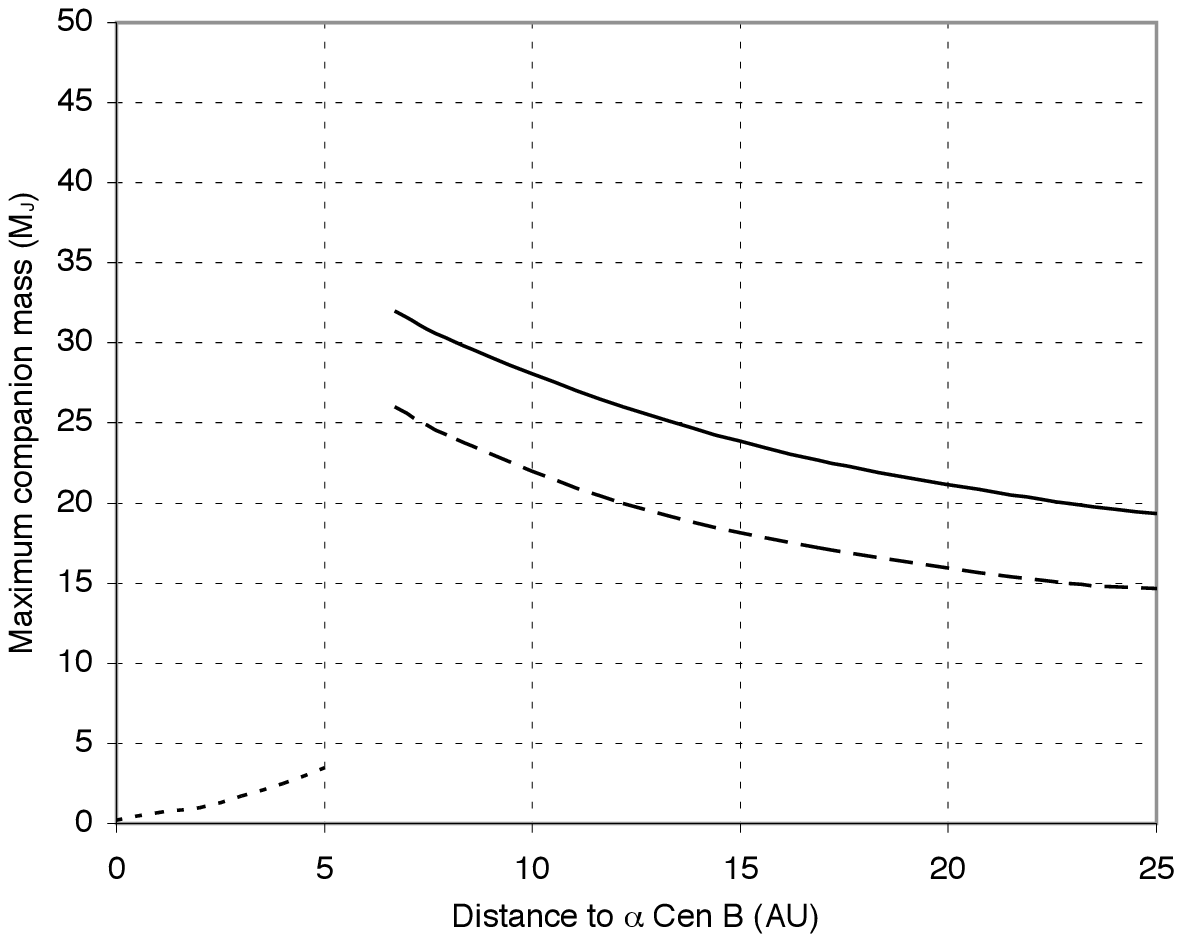}
\caption{Maximum mass of possible companions to $\alpha$\,Cen~B within the
area explored by NACO imaging (encircled zone in Fig.\,\ref{epochs_JHK}, bottom). The solid curve
corresponds to the median magnitude of the detected sources (Fig.~\ref{mag_objects}, bottom)
and the dashed line to the faintest detected objects. The dotted curve corresponds to the
mass limit obtained by Endl et al.~(\cite{endl01}) from radial velocity measurements.}
\label{maxi_companions}
\end{figure}

Massive substellar objects, as opposed to terrestrial planets, are detectable at very large
distances from their parent star, as their magnitude is set by their intrinsic emission
rather than by the reflected light.
The age of the $\alpha$\,Cen system is 5\,Gyr, as determined
by Th\'evenin et al.~(\cite{thevenin02}) and confirmed by the interferometric diameters of the
two stars (Kervella et al.~\cite{kervella03}).
Assuming a mass of 30\,$M_{\rm Jup}$, a 5\,Gyr-old giant planet has absolute
magnitudes of $M_{H} = 18$ and $M_{K} = 20$, from evolutionary models by Baraffe et et al.~(\cite{baraffe03}).
At the distance of $\alpha$\,Cen (1.3\,pc), this translates into apparent
magnitudes $m_H = 16$ and $m_K = 18$, which were within reach of our NACO
imaging search down to an angular separation of $\approx 5$" from $\alpha$\,Cen\,B (Fig.\,\ref{mag_objects}).
Figure~\ref{maxi_companions} gives the limiting sensitivity of our search in terms of
companion mass, based on model magnitudes in the $K$ band from
Baraffe et et al.~(\cite{baraffe03}), and the $K_s$ median magnitudes
given in Fig.~\ref{mag_objects} (bottom). These are conservative estimates,
considering that many sources that are fainter by up to two magnitudes
have been detected in our images.

Murdoch et al.~(\cite{murdoch93}) searched $\alpha$\,Cen\,A and B
for the radial velocity signature of BD companions with orbital
periods $P \le 4000$\,days, but with a negative result.
Using improved measurements obtained over a period of 5.5\,years,
Endl et al.~(\cite{endl01}) concluded that no planet more massive than a few
Jupiter masses, in projected $m \sin i$ value, is orbiting either $\alpha$\,Cen\,A or B within 4\,AU.
If we follow the conclusions of Hale~(\cite{hale94}) that the equatorial planes
of A and B are probably coplanar with the binary orbit plane, and if we accept the hypothesis that
exoplanets orbit in the equatorial plane of their parent star, then this projected mass
value becomes a solid mass limit. The $J$ band search with the
{\it HST} by Schroeder et al.~(\cite{schroeder00}), which did not detect any companion,
was limited to a sensitivity of $m_J = 16$, corresponding to 40\,$M_J$.
Note however that the third star of the $\alpha$\,Cen system {\it Proxima} probably does not
host giant planets. This and other low mass stars were extensively scrutinized
for any radial velocity variation, but did not show any (K\"urster et al.~\cite{kurster03}).
Moreover, speckle-interferometry and imaging surveys (Leinert et al.~\cite{leinert97};
Oppenheimer et al.~\cite{oppenheimer01}) failed to identify
companions down to the BD masses around several low mass stars.
Within our sensitivity and coverage limitations, our negative result leads toward
the modeling results of Wiegert \& Holman~(\cite{wiegert97}), who
conclude that stable companion orbits may not exist beyond about 3\,AU
from each component of the $\alpha$\,Cen pair.

\section{Conclusion}

We have obtained deep adaptive-optics images of the close environment
of $\alpha$\,Cen~A and B. From these images, we did not identify any
comoving companion, but we assembled a catalogue of 252 faint background
objects. Within the explored area, this negative result sets an upper
mass limit of 20-30\,$M_J$ to the possible substellar companions orbiting $\alpha$\,Cen\,B.
If companions of $\alpha$\,Cen~B exist, they are likely to orbit close to the star (within 5 AU)
and to be less massive than a few times Jupiter (from radial velocity surveys). They could
also be fainter than the current imaging search-limiting magnitude, but from Baraffe et al.~(\cite{baraffe03}),
a 5\,Gyr-old, intermediate-mass exoplanet (5\,$M_J$) around $\alpha$\,Cen has an apparent magnitude
of $m_H \approx 27$, and $m_K \approx 39$, fully out of reach of the deepest imaging searches.
Stars younger than $\alpha$\,Cen could be more favorable targets, as the brightness of massive
exoplanets is predicted to decrease steeply with time.
However, the faintness of the detected background sources confirms the capabilities of
modern adaptive optics instruments like NACO for exploring the close environment of
very bright stars and for searching for massive exoplanets. 
\begin{acknowledgements}
Based on observations made with the ESO Telescopes at Cerro Paranal
Observatory under the Director's Discretionary Time (DDT) programs
272.C-5010, 273.C-5041 and 275.C-5027.
We are grateful to ESO's Director General Dr. C. Cesarsky for this
generous allocation of telescope time.
We also wish to thank F. Namouni for fruitful discussions.
This work made use of
observations obtained with the NASA/ESA Hubble Space Telescope,
obtained from the data archive at the Space Telescope Institute.
STScI is operated by the Association of Universities for Research
in Astronomy, Inc. under the NASA contract  NAS 5-26555.
This research made use of the SIMBAD and VIZIER databases
at the CDS, Strasbourg (France), and of NASA's Astrophysics Data
System Bibliographic Services.
\end{acknowledgements}

{}

\begin{scriptsize}
\begin{longtable}{cccccc}
\caption{Position and photometry of the sources detected
around $\alpha$\,Cen. The coordinates are for J2004.5 and
refer to the ICRS. \label{naco_objects1}}\\
\hline \hline
 & $\alpha$ (h:m:s) & $\delta$ ($^\circ$:':") & $m_{\rm J}$ & $m_{\rm H}$ & $m_{\rm Ks}$ \\ 
\hline
\endfirsthead
\caption{continued.}\\
\hline \hline
\# & $\alpha$ (h:m:s) & $\delta$ ($^\circ$:':") & $m_{\rm J}$ & $m_{\rm H}$ & $m_{\rm Ks}$ \\ 
\hline
\endhead
\hline
1 & 14:39:30.257 & -60:50:12.43 & $    $ & $ 20.2 \pm { 0.5 } $ & $ 18.6 \pm { 0.5 } $ \\
2 & 14:39:30.375 & -60:50:03.19 & $    $ & $ 19.4 \pm { 0.7 } $ & $ 20.4 \pm { 1.1 } $ \\
3 & 14:39:30.442 & -60:50:02.73 & $    $ & $ 17.9 \pm { 0.5 } $ & $ 19.6 \pm { 0.6 } $ \\
4 & 14:39:30.511 & -60:50:09.16 & $    $ & $    $ & $ 17.6 \pm { 0.5 } $ \\
5 & 14:39:30.531 & -60:50:11.63 & $    $ & $    $ & $ 19.5 \pm { 0.5 } $ \\
6 & 14:39:30.560 & -60:50:11.80 & $    $ & $    $ & $ 20.4 \pm { 0.6 } $ \\
7 & 14:39:30.650 & -60:50:10.84 & $    $ & $ 18.6 \pm { 0.5 } $ & $ 17.6 \pm { 0.5 } $ \\
8 & 14:39:30.658 & -60:50:13.15 & $    $ & $ 19.4 \pm { 0.5 } $ & $ 18.4 \pm { 0.5 } $ \\
9 & 14:39:30.705 & -60:50:12.36 & $    $ & $    $ & $ 20.7 \pm { 0.7 } $ \\
10 & 14:39:30.727 & -60:50:11.71 & $    $ & $    $ & $ 20.4 \pm { 0.6 } $ \\
11 & 14:39:30.747 & -60:50:02.05 & $    $ & $    $ & $ 18.8 \pm { 0.5 } $ \\
12 & 14:39:30.857 & -60:50:03.98 & $    $ & $ 19.7 \pm { 0.7 } $ & $ 17.7 \pm { 0.5 } $ \\
13 & 14:39:30.935 & -60:50:10.22 & $    $ & $ 18.7 \pm { 0.5 } $ & $ 18.9 \pm { 0.5 } $ \\
14 & 14:39:30.973 & -60:50:06.87 & $    $ & $    $ & $ 15.5 \pm { 0.5 } $ \\
15 & 14:39:30.982 & -60:50:09.65 & $    $ & $ 18.3 \pm { 0.6 } $ & $ 17.3 \pm { 0.5 } $ \\
16 & 14:39:31.024 & -60:50:07.15 & $    $ & $ 17.6 \pm { 0.5 } $ & $ 15.5 \pm { 0.5 } $ \\
17 & 14:39:31.046 & -60:50:05.55 & $    $ & $ 17.6 \pm { 0.5 } $ & $ 17.5 \pm { 0.5 } $ \\
18 & 14:39:31.164 & -60:49:54.27 & $    $ & $    $ & $ 17.1 \pm { 0.5 } $ \\
19 & 14:39:31.166 & -60:50:08.27 & $    $ & $ 17.2 \pm { 0.5 } $ & $ 13.6 \pm { 0.5 } $ \\
20 & 14:39:31.247 & -60:50:06.23 & $ 18.3 \pm { 0.8 } $ & $ 17.5 \pm { 0.5 } $ & $ 18.2 \pm { 0.5 } $ \\
21 & 14:39:31.249 & -60:50:03.54 & $ 18.4 \pm { 0.8 } $ & $ 17.4 \pm { 0.5 } $ & $ 18.3 \pm { 0.5 } $ \\
22 & 14:39:31.253 & -60:50:13.74 & $ 16.7 \pm { 0.8 } $ & $ 16.6 \pm { 0.5 } $ & $ 16.3 \pm { 0.5 } $ \\
23 & 14:39:31.266 & -60:49:53.41 & $    $ & $    $ & $ 19.2 \pm { 0.5 } $ \\
24 & 14:39:31.269 & -60:50:13.63 & $ 17.1 \pm { 0.8 } $ & $ 18.8 \pm { 0.5 } $ & $ 17.6 \pm { 0.5 } $ \\
25 & 14:39:31.296 & -60:50:14.85 & $ 19.2 \pm { 0.8 } $ & $ 19.1 \pm { 0.5 } $ & $ 19.8 \pm { 0.5 } $ \\
26 & 14:39:31.296 & -60:50:02.54 & $ 16.2 \pm { 0.8 } $ & $ 17.2 \pm { 0.5 } $ & $ 18.2 \pm { 0.5 } $ \\
27 & 14:39:31.304 & -60:50:11.06 & $ 15.0 \pm { 0.8 } $ & $ 14.2 \pm { 0.5 } $ & $ 13.4 \pm { 0.5 } $ \\
28 & 14:39:31.309 & -60:50:05.24 & $ 18.1 \pm { 0.8 } $ & $ 17.9 \pm { 0.5 } $ & $ 18.2 \pm { 0.5 } $ \\
29 & 14:39:31.313 & -60:50:09.55 & $   $ & $ 17.3 \pm { 0.5 } $ & $ 16.3 \pm { 0.5 } $ \\
30 & 14:39:31.331 & -60:50:15.18 & $ 18.5 \pm { 0.8 } $ & $ 18.6 \pm { 0.5 } $ & $ 18.7 \pm { 0.5 } $ \\
31 & 14:39:31.334 & -60:50:03.68 & $ 16.5 \pm { 0.8 } $ & $ 16.4 \pm { 0.5 } $ & $ 18.4 \pm { 0.5 } $ \\
32 & 14:39:31.335 & -60:49:55.34 & $    $ & $    $ & $ 18.0 \pm { 0.5 } $ \\
33 & 14:39:31.349 & -60:50:07.94 & $ 14.2 \pm { 0.8 } $ & $ 16.8 \pm { 0.5 } $ & $ 18.0 \pm { 0.5 } $ \\
34 & 14:39:31.376 & -60:50:09.47 & $  $ & $ 17.2 \pm { 0.6 } $ & $ 18.1 \pm { 0.5 } $ \\
35 & 14:39:31.394 & -60:50:08.96 & $ 12.1 \pm { 0.8 } $ & $ 13.1 \pm { 0.5 } $ & $ 13.4 \pm { 0.5 } $ \\
36 & 14:39:31.398 & -60:50:03.96 & $    $ & $ 19.1 \pm { 0.5 } $ & $ 18.5 \pm { 0.5 } $ \\
37 & 14:39:31.407 & -60:49:58.04 & $    $ & $    $ & $ 20.8 \pm { 1.6 } $ \\
38 & 14:39:31.436 & -60:50:14.15 & $ 20.1 \pm { 0.9 } $ & $ 18.3 \pm { 0.5 } $ & $ 18.4 \pm { 0.5 } $ \\
39 & 14:39:31.456 & -60:50:02.46 & $    $ & $ 16.2 \pm { 0.5 } $ & $ 17.3 \pm { 0.5 } $ \\
40 & 14:39:31.461 & -60:50:10.91 & $    $ & $    $ & $ 17.9 \pm { 0.5 } $ \\
41 & 14:39:31.469 & -60:50:04.00 & $ 17.5 \pm { 0.8 } $ & $ 16.7 \pm { 0.5 } $ & $ 17.5 \pm { 0.5 } $ \\
42 & 14:39:31.492 & -60:50:02.44 & $ 16.1 \pm { 0.8 } $ & $ 16.3 \pm { 0.5 } $ & $ 18.7 \pm { 0.5 } $ \\
43 & 14:39:31.509 & -60:50:09.13 & $ 12.7 \pm { 0.8 } $ & $ 16.7 \pm { 0.5 } $ & $ 15.6 \pm { 0.5 } $ \\
44 & 14:39:31.536 & -60:50:15.01 & $    $ & $ 19.7 \pm { 0.5 } $ & $ 20.5 \pm { 0.5 } $ \\
45 & 14:39:31.550 & -60:49:53.86 & $    $ & $    $ & $    $ \\
46 & 14:39:31.599 & -60:50:04.59 & $ 16.7 \pm { 0.8 } $ & $ 16.4 \pm { 0.5 } $ & $ 16.5 \pm { 0.5 } $ \\
47 & 14:39:31.614 & -60:49:55.21 & $    $ & $    $ & $    $ \\
48 & 14:39:31.674 & -60:50:12.89 & $ 17.0 \pm { 0.8 } $ & $ 19.6 \pm { 0.5 } $ & $ 18.7 \pm { 0.5 } $ \\
49 & 14:39:31.679 & -60:50:03.02 & $ 16.5 \pm { 0.8 } $ & $ 17.6 \pm { 0.5 } $ & $ 17.7 \pm { 0.5 } $ \\
50 & 14:39:31.690 & -60:50:04.19 & $ 15.4 \pm { 0.8 } $ & $ 20.4 \pm { 1.0 } $ & $ 16.6 \pm { 0.5 } $ \\
51 & 14:39:31.697 & -60:50:06.79 & $ 15.7 \pm { 0.8 } $ & $    $ & $ 17.6 \pm { 0.5 } $ \\
52 & 14:39:31.701 & -60:50:11.55 & $ 16.4 \pm { 0.8 } $ & $ 18.3 \pm { 0.5 } $ & $ 17.6 \pm { 0.5 } $ \\
53 & 14:39:31.730 & -60:50:03.09 & $ 17.2 \pm { 0.8 } $ & $ 18.0 \pm { 0.5 } $ & $ 17.5 \pm { 0.5 } $ \\
54 & 14:39:31.741 & -60:50:14.40 & $ 17.7 \pm { 0.8 } $ & $ 18.8 \pm { 0.5 } $ & $ 19.0 \pm { 0.5 } $ \\
55 & 14:39:31.788 & -60:49:56.54 & $    $ & $    $ & $    $ \\
56 & 14:39:31.793 & -60:50:15.38 & $ 19.1 \pm { 0.8 } $ & $ 18.4 \pm { 0.5 } $ & $ 19.6 \pm { 0.5 } $ \\
57 & 14:39:31.835 & -60:50:13.58 & $ 18.1 \pm { 0.8 } $ & $ 18.8 \pm { 0.5 } $ & $ 18.7 \pm { 0.5 } $ \\
58 & 14:39:31.842 & -60:50:09.19 & $ 15.3 \pm { 0.8 } $ & $ 14.7 \pm { 0.5 } $ & $ 16.1 \pm { 0.5 } $ \\
59 & 14:39:31.888 & -60:50:08.63 & $ 17.0 \pm { 0.8 } $ & $ 13.8 \pm { 0.5 } $ & $ 17.8 \pm { 0.5 } $ \\
60 & 14:39:31.906 & -60:49:54.64 & $    $ & $    $ & $    $ \\
61 & 14:39:31.960 & -60:50:10.06 & $ 15.6 \pm { 0.8 } $ & $ 16.1 \pm { 0.6 } $ & $ 16.6 \pm { 0.5 } $ \\
62 & 14:39:31.989 & -60:50:17.53 & $    $ & $    $ & $ 18.1 \pm { 0.5 } $ \\
63 & 14:39:31.997 & -60:50:08.50 & $ 17.7 \pm { 1.0 } $ & $ 13.2 \pm { 0.5 } $ & $ 16.4 \pm { 0.5 } $ \\
64 & 14:39:32.029 & -60:50:13.35 & $ 16.5 \pm { 0.8 } $ & $ 19.3 \pm { 0.6 } $ & $ 18.1 \pm { 0.5 } $ \\
65 & 14:39:32.049 & -60:50:17.08 & $    $ & $    $ & $ 13.4 \pm { 0.5 } $ \\
66 & 14:39:32.084 & -60:50:05.95 & $    $ & $ 17.6 \pm { 0.5 } $ & $ 18.5 \pm { 0.5 } $ \\
67 & 14:39:32.093 & -60:50:09.17 & $    $ & $ 17.1 \pm { 0.5 } $ & $ 17.4 \pm { 0.5 } $ \\
68 & 14:39:32.095 & -60:49:57.70 & $    $ & $    $ & $ 17.7 \pm { 0.5 } $ \\
69 & 14:39:32.100 & -60:50:16.77 & $    $ & $    $ & $ 20.1 \pm { 0.7 } $ \\
70 & 14:39:32.105 & -60:50:13.34 & $ 17.1 \pm { 0.8 } $ & $ 19.7 \pm { 0.6 } $ & $ 18.4 \pm { 0.5 } $ \\
71 & 14:39:32.122 & -60:50:24.82 & $    $ & $    $ & $    $ \\
72 & 14:39:32.132 & -60:50:18.03 & $    $ & $    $ & $ 20.4 \pm { 0.7 } $ \\
73 & 14:39:32.154 & -60:50:09.65 & $    $ & $ 16.8 \pm { 0.5 } $ & $ 17.3 \pm { 0.5 } $ \\
74 & 14:39:32.160 & -60:50:17.06 & $    $ & $    $ & $ 20.7 \pm { 0.8 } $ \\
75 & 14:39:32.165 & -60:50:21.96 & $    $ & $    $ & $ 21.2 \pm { 0.6 } $ \\
76 & 14:39:32.178 & -60:50:10.91 & $ 15.9 \pm { 0.8 } $ & $ 17.0 \pm { 0.5 } $ & $ 15.8 \pm { 0.5 } $ \\
77 & 14:39:32.191 & -60:49:57.41 & $    $ & $    $ & $    $ \\
78 & 14:39:32.196 & -60:50:16.04 & $ 17.7 \pm { 0.8 } $ & $    $ & $ 20.2 \pm { 0.9 } $ \\
79 & 14:39:32.203 & -60:50:19.80 & $    $ & $    $ & $ 19.5 \pm { 0.5 } $ \\
80 & 14:39:32.222 & -60:49:54.39 & $    $ & $    $ & $ 19.4 \pm { 0.8 } $ \\
81 & 14:39:32.256 & -60:50:19.74 & $    $ & $    $ & $ 19.6 \pm { 0.5 } $ \\
82 & 14:39:32.276 & -60:50:17.89 & $    $ & $    $ & $ 16.9 \pm { 0.5 } $ \\
83 & 14:39:32.276 & -60:50:02.93 & $    $ & $    $ & $ 12.8 \pm { 0.5 } $ \\
84 & 14:39:32.285 & -60:50:22.85 & $    $ & $    $ & $ 21.5 \pm { 0.9 } $ \\
85 & 14:39:32.329 & -60:49:58.21 & $    $ & $    $ & $ 18.8 \pm { 0.5 } $ \\
86 & 14:39:32.330 & -60:50:24.11 & $    $ & $    $ & $    $ \\
87 & 14:39:32.379 & -60:50:16.73 & $    $ & $    $ & $ 18.8 \pm { 0.5 } $ \\
88 & 14:39:32.381 & -60:50:10.99 & $ 15.2 \pm { 0.8 } $ & $    $ & $ 16.1 \pm { 0.6 } $ \\
89 & 14:39:32.401 & -60:50:18.66 & $    $ & $    $ & $ 18.9 \pm { 0.5 } $ \\
90 & 14:39:32.419 & -60:50:01.17 & $ 13.8 \pm { 0.8 } $ & $    $ & $ 16.2 \pm { 0.5 } $ \\
91 & 14:39:32.439 & -60:50:17.77 & $    $ & $    $ & $ 16.1 \pm { 0.5 } $ \\
92 & 14:39:32.441 & -60:49:55.03 & $ 17.0 \pm { 0.8 } $ & $ 17.2 \pm { 0.5 } $ & $ 18.5 \pm { 0.7 } $ \\
93 & 14:39:32.459 & -60:50:23.06 & $    $ & $ 19.6 \pm { 0.5 } $ & $ 20.0 \pm { 0.5 } $ \\
94 & 14:39:32.468 & -60:50:21.98 & $ 19.0 \pm { 0.8 } $ & $ 17.7 \pm { 0.5 } $ & $ 17.6 \pm { 0.5 } $ \\
95 & 14:39:32.481 & -60:50:24.86 & $ 19.7 \pm { 0.8 } $ & $ 20.3 \pm { 0.5 } $ & $ 19.2 \pm { 0.5 } $ \\
96 & 14:39:32.484 & -60:50:20.17 & $ 21.2 \pm { 1.8 } $ & $ 20.3 \pm { 0.5 } $ & $ 19.2 \pm { 0.5 } $ \\
97 & 14:39:32.497 & -60:50:15.77 & $ 17.7 \pm { 0.9 } $ & $ 19.9 \pm { 0.6 } $ & $ 19.0 \pm { 0.5 } $ \\
98 & 14:39:32.499 & -60:49:57.24 & $ 17.4 \pm { 0.9 } $ & $ 16.4 \pm { 0.5 } $ & $ 17.3 \pm { 0.5 } $ \\
99 & 14:39:32.512 & -60:49:56.03 & $    $ & $ 16.9 \pm { 0.5 } $ & $ 18.0 \pm { 0.5 } $ \\
100 & 14:39:32.553 & -60:50:09.28 & $ 13.4 \pm { 0.8 } $ & $ 15.4 \pm { 0.5 } $ & $ 17.4 \pm { 0.9 } $ \\
101 & 14:39:32.555 & -60:50:09.43 & $    $ & $    $ & $    $ \\
102 & 14:39:32.561 & -60:50:22.14 & $ 21.3 \pm { 1.1 } $ & $    $ & $ 22.5 \pm { 0.9 } $ \\
103 & 14:39:32.615 & -60:50:16.36 & $ 17.4 \pm { 0.8 } $ & $ 20.2 \pm { 0.9 } $ & $ 18.0 \pm { 0.5 } $ \\
104 & 14:39:32.631 & -60:50:08.17 & $ 12.9 \pm { 0.8 } $ & $ 15.0 \pm { 0.5 } $ & $    $ \\
105 & 14:39:32.640 & -60:50:16.60 & $ 18.1 \pm { 0.9 } $ & $ 18.4 \pm { 0.5 } $ & $ 17.9 \pm { 0.5 } $ \\
106 & 14:39:32.648 & -60:50:17.94 & $    $ & $ 19.9 \pm { 0.6 } $ & $ 19.5 \pm { 0.5 } $ \\
107 & 14:39:32.653 & -60:50:22.34 & $    $ & $ 19.3 \pm { 0.5 } $ & $ 20.3 \pm { 0.5 } $ \\
108 & 14:39:32.686 & -60:50:24.89 & $    $ & $ 20.3 \pm { 0.5 } $ & $ 19.5 \pm { 0.5 } $ \\
109 & 14:39:32.704 & -60:50:22.37 & $    $ & $    $ & $ 21.0 \pm { 0.5 } $ \\
110 & 14:39:32.726 & -60:50:01.05 & $    $ & $ 12.7 \pm { 0.5 } $ & $ 12.7 \pm { 0.5 } $ \\
111 & 14:39:32.729 & -60:50:06.40 & $ 12.5 \pm { 0.8 } $ & $ 14.9 \pm { 0.5 } $ & $ 15.4 \pm { 0.5 } $ \\
112 & 14:39:32.738 & -60:49:52.85 & $    $ & $ 17.7 \pm { 0.5 } $ & $ 17.2 \pm { 0.5 } $ \\
113 & 14:39:32.755 & -60:50:19.36 & $ 19.0 \pm { 0.8 } $ & $ 17.4 \pm { 0.5 } $ & $ 17.3 \pm { 0.5 } $ \\
114 & 14:39:32.771 & -60:50:20.01 & $    $ & $ 20.2 \pm { 0.6 } $ & $ 21.8 \pm { 0.9 } $ \\
115 & 14:39:32.775 & -60:50:22.75 & $ 14.9 \pm { 0.8 } $ & $ 18.9 \pm { 0.5 } $ & $ 17.1 \pm { 0.5 } $ \\
116 & 14:39:32.777 & -60:49:53.95 & $    $ & $ 21.7 \pm { 0.9 } $ & $ 21.3 \pm { 0.6 } $ \\
117 & 14:39:32.789 & -60:50:09.45 & $ 14.4 \pm { 1.0 } $ & $    $ & $    $ \\
118 & 14:39:32.811 & -60:50:19.09 & $ 19.0 \pm { 0.9 } $ & $ 17.1 \pm { 0.5 } $ & $ 16.7 \pm { 0.5 } $ \\
119 & 14:39:32.815 & -60:50:24.17 & $ 18.9 \pm { 0.8 } $ & $ 20.4 \pm { 0.5 } $ & $ 19.0 \pm { 0.5 } $ \\
120 & 14:39:32.816 & -60:50:15.12 & $ 15.7 \pm { 0.8 } $ & $ 17.5 \pm { 0.5 } $ & $ 15.9 \pm { 0.5 } $ \\
121 & 14:39:32.824 & -60:50:09.15 & $ 14.1 \pm { 0.9 } $ & $ 13.5 \pm { 0.5 } $ & $ 14.1 \pm { 0.5 } $ \\
122 & 14:39:32.824 & -60:49:55.56 & $    $ & $ 17.3 \pm { 0.5 } $ & $ 18.1 \pm { 0.6 } $ \\
123 & 14:39:32.840 & -60:50:07.19 & $ 12.5 \pm { 0.8 } $ & $ 14.1 \pm { 0.5 } $ & $    $ \\
124 & 14:39:32.847 & -60:50:20.09 & $    $ & $ 19.8 \pm { 0.5 } $ & $ 20.1 \pm { 0.5 } $ \\
125 & 14:39:32.855 & -60:50:24.07 & $ 18.9 \pm { 0.8 } $ & $ 18.2 \pm { 0.5 } $ & $ 17.9 \pm { 0.5 } $ \\
126 & 14:39:32.887 & -60:50:16.77 & $    $ & $ 18.7 \pm { 0.5 } $ & $ 16.8 \pm { 0.5 } $ \\
127 & 14:39:32.896 & -60:50:24.83 & $ 19.3 \pm { 0.8 } $ & $ 18.9 \pm { 0.5 } $ & $ 18.4 \pm { 0.5 } $ \\
128 & 14:39:32.898 & -60:50:19.76 & $    $ & $    $ & $ 22.3 \pm { 1.7 } $ \\
129 & 14:39:32.913 & -60:50:25.54 & $    $ & $    $ & $ 17.5 \pm { 0.5 } $ \\
130 & 14:39:32.945 & -60:50:10.90 & $    $ & $ 14.2 \pm { 0.5 } $ & $ 11.8 \pm { 0.5 } $ \\
131 & 14:39:32.951 & -60:50:21.44 & $    $ & $    $ & $ 20.6 \pm { 0.5 } $ \\
132 & 14:39:32.954 & -60:50:13.79 & $ 16.6 \pm { 0.9 } $ & $ 16.3 \pm { 0.5 } $ & $ 15.1 \pm { 0.5 } $ \\
133 & 14:39:32.958 & -60:50:23.30 & $ 20.3 \pm { 0.9 } $ & $ 18.5 \pm { 0.5 } $ & $ 18.6 \pm { 0.5 } $ \\
134 & 14:39:32.958 & -60:50:18.80 & $ 19.1 \pm { 1.0 } $ & $ 20.3 \pm { 0.6 } $ & $ 19.3 \pm { 0.5 } $ \\
135 & 14:39:33.014 & -60:50:05.71 & $    $ & $    $ & $ 14.1 \pm { 0.5 } $ \\
136 & 14:39:33.038 & -60:50:15.74 & $ 15.8 \pm { 0.8 } $ & $ 17.3 \pm { 0.5 } $ & $ 18.4 \pm { 0.5 } $ \\
137 & 14:39:33.061 & -60:50:21.57 & $    $ & $ 18.6 \pm { 0.5 } $ & $ 18.6 \pm { 0.5 } $ \\
138 & 14:39:33.092 & -60:49:52.87 & $    $ & $ 20.0 \pm { 1.0 } $ & $ 17.2 \pm { 0.5 } $ \\
139 & 14:39:33.099 & -60:50:18.78 & $ 16.4 \pm { 0.8 } $ & $    $ & $ 18.9 \pm { 0.5 } $ \\
140 & 14:39:33.109 & -60:50:21.29 & $ 18.4 \pm { 0.8 } $ & $    $ & $ 21.5 \pm { 0.8 } $ \\
141 & 14:39:33.114 & -60:50:12.22 & $    $ & $ 15.1 \pm { 0.5 } $ & $ 15.0 \pm { 0.5 } $ \\
142 & 14:39:33.141 & -60:50:19.90 & $ 18.2 \pm { 0.8 } $ & $ 20.2 \pm { 0.7 } $ & $ 21.8 \pm { 0.7 } $ \\
143 & 14:39:33.167 & -60:50:23.69 & $    $ & $ 18.8 \pm { 0.5 } $ & $ 18.2 \pm { 0.5 } $ \\
144 & 14:39:33.179 & -60:49:57.53 & $    $ & $ 18.6 \pm { 2.1 } $ & $ 15.1 \pm { 0.5 } $ \\
145 & 14:39:33.212 & -60:50:18.95 & $ 13.9 \pm { 0.8 } $ & $    $ & $ 18.8 \pm { 0.6 } $ \\
146 & 14:39:33.237 & -60:50:12.29 & $    $ & $ 15.2 \pm { 0.5 } $ & $ 15.4 \pm { 0.5 } $ \\
147 & 14:39:33.265 & -60:50:21.57 & $ 16.8 \pm { 0.8 } $ & $ 19.0 \pm { 0.5 } $ & $ 19.2 \pm { 0.5 } $ \\
148 & 14:39:33.288 & -60:50:17.69 & $ 13.6 \pm { 0.8 } $ & $ 15.1 \pm { 0.5 } $ & $ 16.2 \pm { 0.5 } $ \\
149 & 14:39:33.292 & -60:50:12.37 & $ 13.8 \pm { 0.8 } $ & $ 17.2 \pm { 1.3 } $ & $ 15.6 \pm { 0.5 } $ \\
150 & 14:39:33.317 & -60:49:54.18 & $ 14.3 \pm { 0.8 } $ & $ 15.2 \pm { 0.5 } $ & $    $ \\
151 & 14:39:33.333 & -60:50:00.36 & $ 13.7 \pm { 0.8 } $ & $ 15.4 \pm { 0.5 } $ & $ 18.6 \pm { 1.0 } $ \\
152 & 14:39:33.353 & -60:50:22.36 & $ 19.3 \pm { 0.8 } $ & $ 19.8 \pm { 0.5 } $ & $ 21.2 \pm { 0.6 } $ \\
153 & 14:39:33.368 & -60:50:12.77 & $ 13.2 \pm { 0.8 } $ & $ 14.6 \pm { 0.5 } $ & $ 16.1 \pm { 0.5 } $ \\
154 & 14:39:33.379 & -60:49:54.96 & $ 14.1 \pm { 0.8 } $ & $ 17.4 \pm { 0.5 } $ & $ 16.9 \pm { 0.5 } $ \\
155 & 14:39:33.392 & -60:50:23.52 & $ 19.7 \pm { 0.9 } $ & $ 21.3 \pm { 0.7 } $ & $ 19.9 \pm { 0.5 } $ \\
156 & 14:39:33.444 & -60:50:15.73 & $ 15.3 \pm { 0.8 } $ & $    $ & $ 17.4 \pm { 0.5 } $ \\
157 & 14:39:33.453 & -60:50:17.61 & $ 17.5 \pm { 0.9 } $ & $    $ & $ 18.4 \pm { 0.5 } $ \\
158 & 14:39:33.460 & -60:50:24.45 & $    $ & $ 16.8 \pm { 0.5 } $ & $ 16.3 \pm { 0.5 } $ \\
159 & 14:39:33.486 & -60:50:17.87 & $ 18.4 \pm { 1.1 } $ & $    $ & $ 17.5 \pm { 0.5 } $ \\
160 & 14:39:33.495 & -60:50:15.18 & $ 14.3 \pm { 0.8 } $ & $ 16.8 \pm { 0.5 } $ & $ 17.0 \pm { 0.5 } $ \\
161 & 14:39:33.502 & -60:50:16.15 & $ 16.1 \pm { 0.8 } $ & $ 17.0 \pm { 0.5 } $ & $ 17.7 \pm { 0.5 } $ \\
162 & 14:39:33.506 & -60:50:20.31 & $ 18.4 \pm { 0.8 } $ & $    $ & $ 22.5 \pm { 3.1 } $ \\
163 & 14:39:33.538 & -60:50:18.66 & $ 18.4 \pm { 0.9 } $ & $ 17.0 \pm { 0.5 } $ & $ 17.0 \pm { 0.5 } $ \\
164 & 14:39:33.551 & -60:50:25.31 & $ 18.0 \pm { 0.8 } $ & $ 18.8 \pm { 0.5 } $ & $ 17.8 \pm { 0.5 } $ \\
165 & 14:39:33.555 & -60:50:18.03 & $ 17.8 \pm { 0.9 } $ & $ 18.7 \pm { 0.5 } $ & $ 19.1 \pm { 0.5 } $ \\
166 & 14:39:33.562 & -60:50:23.63 & $    $ & $ 20.8 \pm { 0.6 } $ & $ 19.1 \pm { 0.5 } $ \\
167 & 14:39:33.567 & -60:50:11.57 & $ 13.0 \pm { 0.8 } $ & $ 12.3 \pm { 0.5 } $ & $ 12.1 \pm { 0.5 } $ \\
168 & 14:39:33.571 & -60:50:20.84 & $ 19.6 \pm { 0.9 } $ & $ 20.9 \pm { 0.9 } $ & $ 19.8 \pm { 0.5 } $ \\
169 & 14:39:33.591 & -60:50:13.11 & $ 13.9 \pm { 0.8 } $ & $ 15.6 \pm { 0.5 } $ & $ 15.7 \pm { 0.5 } $ \\
170 & 14:39:33.606 & -60:50:24.01 & $ 17.8 \pm { 0.8 } $ & $ 16.9 \pm { 0.5 } $ & $ 16.9 \pm { 0.5 } $ \\
171 & 14:39:33.624 & -60:50:17.95 & $ 19.1 \pm { 1.4 } $ & $ 19.1 \pm { 0.5 } $ & $    $ \\
172 & 14:39:33.665 & -60:49:53.58 & $ 15.5 \pm { 0.8 } $ & $ 16.0 \pm { 0.5 } $ & $ 16.7 \pm { 0.5 } $ \\
173 & 14:39:33.669 & -60:50:16.53 & $ 17.6 \pm { 0.8 } $ & $ 18.7 \pm { 0.5 } $ & $ 19.0 \pm { 0.5 } $ \\
174 & 14:39:33.676 & -60:50:19.72 & $ 17.7 \pm { 0.8 } $ & $ 19.0 \pm { 0.5 } $ & $ 20.4 \pm { 0.5 } $ \\
175 & 14:39:33.680 & -60:50:18.78 & $ 20.6 \pm { 1.1 } $ & $    $ & $ 19.6 \pm { 0.5 } $ \\
176 & 14:39:33.687 & -60:50:17.54 & $    $ & $ 18.2 \pm { 0.5 } $ & $ 18.3 \pm { 0.5 } $ \\
177 & 14:39:33.696 & -60:50:17.01 & $ 18.2 \pm { 0.8 } $ & $ 17.8 \pm { 0.5 } $ & $ 17.8 \pm { 0.5 } $ \\
178 & 14:39:33.707 & -60:50:05.00 & $ 14.5 \pm { 0.8 } $ & $ 16.1 \pm { 0.5 } $ & $ 17.9 \pm { 0.9 } $ \\
179 & 14:39:33.709 & -60:50:19.01 & $ 18.9 \pm { 1.0 } $ & $ 21.2 \pm { 1.4 } $ & $ 19.2 \pm { 0.5 } $ \\
180 & 14:39:33.713 & -60:50:13.35 & $ 14.9 \pm { 0.8 } $ & $ 16.2 \pm { 0.5 } $ & $ 17.4 \pm { 0.5 } $ \\
181 & 14:39:33.725 & -60:50:24.73 & $ 18.8 \pm { 0.8 } $ & $    $ & $ 19.1 \pm { 0.5 } $ \\
182 & 14:39:33.727 & -60:50:20.64 & $    $ & $ 19.6 \pm { 0.5 } $ & $ 19.1 \pm { 0.5 } $ \\
183 & 14:39:33.785 & -60:50:20.70 & $ 18.4 \pm { 0.8 } $ & $ 18.4 \pm { 0.5 } $ & $ 18.2 \pm { 0.5 } $ \\
184 & 14:39:33.803 & -60:50:19.48 & $ 17.4 \pm { 0.8 } $ & $    $ & $    $ \\
185 & 14:39:33.811 & -60:50:17.41 & $ 17.7 \pm { 0.8 } $ & $ 19.0 \pm { 0.5 } $ & $ 19.3 \pm { 0.5 } $ \\
186 & 14:39:33.831 & -60:50:13.80 & $    $ & $ 18.2 \pm { 0.5 } $ & $ 17.8 \pm { 0.5 } $ \\
187 & 14:39:33.836 & -60:49:58.82 & $ 11.9 \pm { 0.8 } $ & $ 13.3 \pm { 0.5 } $ & $    $ \\
188 & 14:39:33.841 & -60:50:05.90 & $    $ & $ 16.7 \pm { 0.7 } $ & $ 15.9 \pm { 0.5 } $ \\
189 & 14:39:33.845 & -60:50:18.84 & $    $ & $    $ & $ 22.2 \pm { 1.9 } $ \\
190 & 14:39:33.861 & -60:50:08.55 & $    $ & $ 15.9 \pm { 0.5 } $ & $ 14.1 \pm { 0.5 } $ \\
191 & 14:39:33.868 & -60:49:55.04 & $ 14.6 \pm { 0.8 } $ & $ 16.7 \pm { 1.2 } $ & $    $ \\
192 & 14:39:33.876 & -60:50:11.32 & $ 14.9 \pm { 0.8 } $ & $ 16.6 \pm { 0.5 } $ & $ 16.6 \pm { 0.5 } $ \\
193 & 14:39:33.909 & -60:50:13.59 & $ 15.9 \pm { 0.8 } $ & $ 17.4 \pm { 0.5 } $ & $ 17.7 \pm { 0.5 } $ \\
194 & 14:39:33.910 & -60:50:17.24 & $ 17.8 \pm { 0.8 } $ & $ 18.9 \pm { 0.5 } $ & $ 17.8 \pm { 0.5 } $ \\
195 & 14:39:33.919 & -60:50:19.77 & $ 18.1 \pm { 0.8 } $ & $    $ & $ 19.5 \pm { 0.5 } $ \\
196 & 14:39:33.945 & -60:50:21.36 & $    $ & $    $ & $ 20.1 \pm { 0.5 } $ \\
197 & 14:39:33.956 & -60:50:03.90 & $    $ & $    $ & $    $ \\
198 & 14:39:33.965 & -60:49:54.50 & $    $ & $ 16.0 \pm { 0.6 } $ & $    $ \\
199 & 14:39:33.968 & -60:49:58.86 & $    $ & $ 13.7 \pm { 0.5 } $ & $ 14.2 \pm { 0.5 } $ \\
200 & 14:39:33.976 & -60:50:15.80 & $ 16.8 \pm { 0.8 } $ & $ 18.5 \pm { 0.5 } $ & $ 18.8 \pm { 0.5 } $ \\
201 & 14:39:33.999 & -60:50:22.55 & $    $ & $ 21.8 \pm { 0.9 } $ & $ 20.4 \pm { 0.5 } $ \\
202 & 14:39:34.005 & -60:50:17.33 & $ 16.2 \pm { 0.8 } $ & $ 18.0 \pm { 0.5 } $ & $ 20.6 \pm { 0.6 } $ \\
203 & 14:39:34.006 & -60:50:16.59 & $ 15.6 \pm { 0.8 } $ & $ 17.7 \pm { 0.5 } $ & $ 18.3 \pm { 0.5 } $ \\
204 & 14:39:34.028 & -60:50:25.17 & $    $ & $    $ & $    $ \\
205 & 14:39:34.037 & -60:50:19.77 & $    $ & $ 19.8 \pm { 0.6 } $ & $ 19.1 \pm { 0.5 } $ \\
206 & 14:39:34.068 & -60:50:23.75 & $    $ & $ 19.5 \pm { 0.5 } $ & $ 19.7 \pm { 0.5 } $ \\
207 & 14:39:34.077 & -60:50:17.89 & $ 17.2 \pm { 0.8 } $ & $ 19.2 \pm { 0.5 } $ & $ 19.0 \pm { 0.5 } $ \\
208 & 14:39:34.079 & -60:49:53.16 & $ 15.2 \pm { 0.8 } $ & $ 15.0 \pm { 0.5 } $ & $ 16.1 \pm { 0.5 } $ \\
209 & 14:39:34.083 & -60:50:15.38 & $ 17.4 \pm { 0.8 } $ & $ 19.1 \pm { 0.5 } $ & $ 18.8 \pm { 0.5 } $ \\
210 & 14:39:34.083 & -60:50:14.19 & $ 17.0 \pm { 0.8 } $ & $    $ & $ 18.1 \pm { 0.5 } $ \\
211 & 14:39:34.101 & -60:50:16.77 & $ 17.9 \pm { 0.8 } $ & $ 18.1 \pm { 0.5 } $ & $ 17.6 \pm { 0.5 } $ \\
212 & 14:39:34.144 & -60:50:24.99 & $ 19.3 \pm { 0.8 } $ & $    $ & $    $ \\
213 & 14:39:34.152 & -60:50:17.32 & $ 17.0 \pm { 0.8 } $ & $ 16.6 \pm { 0.5 } $ & $ 16.5 \pm { 0.5 } $ \\
214 & 14:39:34.153 & -60:50:20.33 & $ 18.7 \pm { 0.8 } $ & $ 20.5 \pm { 1.0 } $ & $ 19.2 \pm { 0.5 } $ \\
215 & 14:39:34.155 & -60:50:23.22 & $    $ & $ 20.0 \pm { 0.5 } $ & $ 20.5 \pm { 0.6 } $ \\
216 & 14:39:34.163 & -60:50:24.03 & $    $ & $ 19.6 \pm { 0.5 } $ & $ 18.8 \pm { 0.5 } $ \\
217 & 14:39:34.168 & -60:50:15.10 & $ 17.9 \pm { 0.8 } $ & $ 19.2 \pm { 0.5 } $ & $ 17.3 \pm { 0.5 } $ \\
218 & 14:39:34.175 & -60:50:21.65 & $ 17.9 \pm { 0.8 } $ & $ 18.1 \pm { 0.5 } $ & $ 17.0 \pm { 0.5 } $ \\
219 & 14:39:34.181 & -60:50:21.76 & $ 17.8 \pm { 0.8 } $ & $ 18.5 \pm { 0.5 } $ & $ 17.0 \pm { 0.5 } $ \\
220 & 14:39:34.182 & -60:50:19.49 & $ 16.6 \pm { 0.8 } $ & $    $ & $ 19.3 \pm { 0.5 } $ \\
221 & 14:39:34.191 & -60:50:11.60 & $ 14.5 \pm { 0.8 } $ & $ 18.3 \pm { 0.8 } $ & $ 18.8 \pm { 0.7 } $ \\
222 & 14:39:34.193 & -60:50:16.60 & $ 18.9 \pm { 0.8 } $ & $ 18.7 \pm { 0.5 } $ & $ 17.4 \pm { 0.5 } $ \\
223 & 14:39:34.235 & -60:50:16.35 & $ 17.3 \pm { 0.8 } $ & $ 18.7 \pm { 0.5 } $ & $ 18.2 \pm { 0.5 } $ \\
224 & 14:39:34.290 & -60:50:16.34 & $    $ & $    $ & $ 16.9 \pm { 0.5 } $ \\
225 & 14:39:34.322 & -60:50:15.95 & $    $ & $ 17.9 \pm { 0.5 } $ & $ 17.4 \pm { 0.5 } $ \\
226 & 14:39:34.389 & -60:50:14.93 & $ 17.8 \pm { 0.8 } $ & $ 18.9 \pm { 0.5 } $ & $ 19.0 \pm { 0.5 } $ \\
227 & 14:39:34.476 & -60:50:14.77 & $ 19.1 \pm { 1.0 } $ & $ 20.7 \pm { 1.1 } $ & $ 19.7 \pm { 0.6 } $ \\
228 & 14:39:34.538 & -60:50:14.05 & $ 17.4 \pm { 0.8 } $ & $ 20.0 \pm { 0.8 } $ & $ 17.4 \pm { 0.5 } $ \\
229 & 14:39:34.616 & -60:50:11.46 & $ 15.5 \pm { 0.8 } $ & $ 17.7 \pm { 0.5 } $ & $ 17.1 \pm { 0.5 } $ \\
230 & 14:39:34.631 & -60:50:17.45 & $    $ & $ 18.8 \pm { 0.5 } $ & $ 19.6 \pm { 0.5 } $ \\
231 & 14:39:34.634 & -60:50:13.18 & $    $ & $ 20.6 \pm { 1.5 } $ & $ 18.1 \pm { 0.6 } $ \\
232 & 14:39:34.661 & -60:50:17.75 & $ 17.6 \pm { 0.8 } $ & $    $ & $ 20.5 \pm { 0.6 } $ \\
233 & 14:39:34.709 & -60:50:16.04 & $    $ & $ 21.5 \pm { 1.7 } $ & $ 16.7 \pm { 0.5 } $ \\
234 & 14:39:34.732 & -60:50:16.97 & $    $ & $ 20.8 \pm { 0.8 } $ & $ 17.5 \pm { 0.5 } $ \\
235 & 14:39:34.826 & -60:50:16.65 & $ 18.7 \pm { 0.9 } $ & $    $ & $ 22.2 \pm { 1.4 } $ \\
236 & 14:39:34.906 & -60:50:07.49 & $ 15.0 \pm { 0.8 } $ & $ 16.4 \pm { 0.5 } $ & $ 17.0 \pm { 0.5 } $ \\
237 & 14:39:34.928 & -60:50:12.36 & $    $ & $    $ & $ 20.0 \pm { 0.6 } $ \\
238 & 14:39:34.941 & -60:50:14.45 & $ 18.7 \pm { 0.9 } $ & $ 21.4 \pm { 1.3 } $ & $ 21.9 \pm { 1.2 } $ \\
239 & 14:39:35.022 & -60:50:16.89 & $    $ & $ 18.8 \pm { 0.5 } $ & $ 19.1 \pm { 0.5 } $ \\
240 & 14:39:35.055 & -60:50:14.96 & $ 18.2 \pm { 0.8 } $ & $ 19.9 \pm { 0.6 } $ & $ 18.6 \pm { 0.5 } $ \\
241 & 14:39:35.082 & -60:50:06.27 & $ 14.0 \pm { 0.8 } $ & $ 16.6 \pm { 0.5 } $ & $ 16.7 \pm { 0.5 } $ \\
242 & 14:39:35.124 & -60:50:18.65 & $    $ & $ 18.2 \pm { 0.5 } $ & $ 17.8 \pm { 0.5 } $ \\
243 & 14:39:35.131 & -60:50:09.45 & $ 16.7 \pm { 0.8 } $ & $ 18.1 \pm { 0.5 } $ & $ 18.1 \pm { 0.5 } $ \\
244 & 14:39:35.202 & -60:50:10.35 & $ 16.5 \pm { 0.8 } $ & $ 18.3 \pm { 0.5 } $ & $    $ \\
245 & 14:39:35.224 & -60:50:06.92 & $ 18.4 \pm { 1.2 } $ & $ 20.2 \pm { 1.5 } $ & $ 18.3 \pm { 0.5 } $ \\
246 & 14:39:35.256 & -60:50:12.48 & $    $ & $ 19.9 \pm { 0.6 } $ & $ 19.6 \pm { 0.5 } $ \\
247 & 14:39:35.648 & -60:50:04.41 & $ 17.6 \pm { 1.0 } $ & $ 18.8 \pm { 0.8 } $ & $ 17.2 \pm { 0.5 } $ \\
248 & 14:39:35.682 & -60:49:57.24 & $ 15.9 \pm { 0.8 } $ & $ 16.7 \pm { 0.5 } $ & $ 17.0 \pm { 0.5 } $ \\
249 & 14:39:35.802 & -60:49:57.48 & $    $ & $ 18.0 \pm { 0.6 } $ & $ 17.0 \pm { 0.5 } $ \\
250 & 14:39:35.929 & -60:50:06.15 & $ 18.2 \pm { 0.8 } $ & $ 17.5 \pm { 0.5 } $ & $ 16.7 \pm { 0.5 } $ \\
251 & 14:39:35.956 & -60:50:00.28 & $ 14.7 \pm { 0.8 } $ & $ 15.9 \pm { 0.5 } $ & $ 17.1 \pm { 0.5 } $ \\
252 & 14:39:36.451 & -60:49:57.58 & $ 16.9 \pm { 0.8 } $ & $ 17.4 \pm { 0.5 } $ & $ 16.7 \pm { 0.5 } $ \\\hline
\end{longtable}
\end{scriptsize}

\end{document}